\documentclass[10pt]{IEEEtran}

\usepackage{algorithmic}
\usepackage{algorithm}
\usepackage{complexity}
\usepackage{times}
\usepackage{graphicx}
\usepackage{multirow}
\usepackage{slashbox}
\usepackage{tabularx}
\newtheorem{theorem}{Theorem}
\newtheorem{lemma}[theorem]{Lemma}

\newtheorem{example}[theorem]{Example}
\newtheorem{definition}{Definition}
\usepackage{cite}
\usepackage{amsmath}
\usepackage{amssymb}
\usepackage{epsfig}
\usepackage{color}

\newcommand{\ldbrac}{[\![}
\newcommand{\rdbrac}{]\!]} 
\newcommand{\ldabrac}{\langle\!\langle}
\newcommand{\rdabrac}{\rangle\!\rangle} 
\newcommand{\emgfp}{\mbox{\bf gfp}} 
\newcommand{\emlfp}{\mbox{\bf lfp}}

\newcommand{\emplay}{\mbox{\em play}}

\newcommand{\emspms}{{\mbox{\em SP}^\calmdl_\calspc}} 
\newcommand{\emepms}{{\mbox{\em EP}^\calmdl_\calspc}} 
\newcommand{\emsepms}{{\mbox{\scriptsize \em EP}^\calmdl_\calspc}} 
\newcommand{\emufms}{{\mbox{\em UF}^\calmdl_\calspc}}

\newcommand{\emass}{\mbox{\em reset}} 
 
\newcommand{\cala}{{\cal A}} 
\newcommand{\calb}{{\cal B}} 
\newcommand{\calab}{{\cal A\times B}} 
\newcommand{\calenv}{{\cal E}} 
\newcommand{\calmdl}{{\cal M}} 
\newcommand{\calspc}{{\cal S}} 
\newcommand{\calemdl}{{\calenv\times\calmdl}} 
\newcommand{\calespc}{{\calenv\times\calspc}}

\newcommand{\true}{\mbox{\em true}}
\newcommand{\false}{\mbox{\em false}}

\newcommand{\pf}{\noindent\mbox{\bf Proof : }}

\newcommand{\defn}{\stackrel{\mbox{\tiny def}}{=}}

\newcommand{\pfrr}{\Box}
\newcommand{\until}{\mbox{$\cal U$}}

\newcommand{\nxt}{\bigcirc}

\newcommand{\ttmode}{\mbox{\tt mode}} 

\newcommand{\bbbbb}{{\mathbb B}}

\newcommand{\rnall}{{\mathbb R}}
\newcommand{\rnneg}{{{\mathbb R}^{\geq 0}}}
\newcommand{\nnneg}{{\mathbb N}}

\newcommand{\bbbbs}{{\mathbb S}}

\newcommand{\ccmfs}{{C^\calmdl_\calspc}}

\newcommand{\llrarrow}{{\rule[2pt]{8mm}{0.5pt}\!\!\!\longrightarrow}}
\newcommand{\lllrarrow}{{\rule[2pt]{10mm}{0.5pt}\!\!\!\longrightarrow}}

\def\qed{\ifmmode\|\else{\unskip\nobreak\hfil
\penalty50\hskip1em\null\nobreak\hfil$\blacksquare$
\parfillskip=0pt\finalhyphendemerits=0\endgraf}\fi}

\newenvironment{list1}{\begin{list}{$\bullet$}
{\topsep 0 pt \parsep 0 pt \partopsep 0 pt \itemsep 0 pt}}{\end{list}}
\newenvironment{list2}{\begin{list}{$-$}
{\topsep 0 pt \parsep 0 pt \partopsep 0 pt \itemsep 0 pt}}{\end{list}}
\newenvironment{list3}{\begin{list}{$*$}
{\topsep 0 pt \parsep 0 pt \partopsep 0 pt \itemsep 0 pt}}{\end{list}}

\newcounter{cabbage1}

\newcounter{cabbage2}

\newcounter{cabbage3}

\newcounter{bean1}

\newcounter{bean2}

\newcounter{bean3}

\newcounter{bean4}

\newcounter{bean5}

\newcounter{bean6}

\begin{document}

\title{Simulation-Checking of Real-Time Systems with 
Fairness Assumptions
}

\author{Farn Wang \\[2mm]
Dept. of Electrical Engineering \&
Graduate Institute of Electronic Engineering \\
National Taiwan University\\
farn@cc.ee.ntu.edu.tw;
http://cc.ee.ntu.edu.tw/\~{ }farn\\[2mm]
{\bf REDLIB} is available at
http://sites.google.com/site/redlibtw/.
}

\maketitle
\thispagestyle{empty}
\pagestyle{plain}

\begin{abstract} 
We investigate the simulation problem in of dense-time system.  
A specification simulates a model if the specification can 
match every transition that the model can make at a time point.  
We also adapt the approach of Emerson and Lei 
and allow for multiple strong and weak fairness assumptions 
in checking the simulation relation.  
Furthermore, we allow for fairness assumptions specified as 
either state-predicates or event-predicates.  
We focus on a subclass of the problem with at most one fairness assumption 
for the specification.  
We then present a simulation-checking algorithm for this subclass.  
We propose simulation of a model by a specification against a 
common environment.  
We present efficient techniques for such simulations to take  
the common environment into consideration.  
Our experiment shows that such a consideration can dramatically 
improve the efficiency of checking simulation.  
We also report the performance of our algorithm in checking the liveness 
properties with fairness assumptions. 
\end{abstract}


\section{Introduction}

Modern {\em real-time systems} have incurred tremendous challenges 
to verification engineers.  
The reason is that 
a model process running in a modern real-time system   
can be built with support from many server processes in the environment.  
Moreover, the model may also have to respond to requests 
from several user processes.  
The fulfillment of a computation 
relies not only on the functional correctness of the model, 
but also on the reactions from the servers and the clients.  
For example, a company may submit a task of DNA sequencing to a server. 
The server then develops a computing budget and 
decomposes the task into several subtasks (e.g., SNP finding, alignments).  
Then the server may relegate the subtasks to several other servers.  
The decompositions of subtasks may then go on and on.  
If the task is to be completed, not only the server for the root task 
needs to function correctly, 
but also all the servers for the subtasks have to fulfill their assignments.  
Thus, to verify the function of the root server,  
it is only reasonable and practical to assume that all the other 
supporting servers work correctly.  

In many industrial projects, 
the specification can be given in the concept of state-transition 
diagrams (or tables).  
In such a context, {\em simulation-checking} is an appropriate framework 
for verifying that a model conforms to the behavior of a specification
\cite{Cerans92,TAKB96}.   
Intuitively, the specification simulates the model 
if every timed step of the model can be matched by the specification 
at the same time.  

\begin{figure*}
\begin{center} 
\begin{picture}(0,0)%
\includegraphics{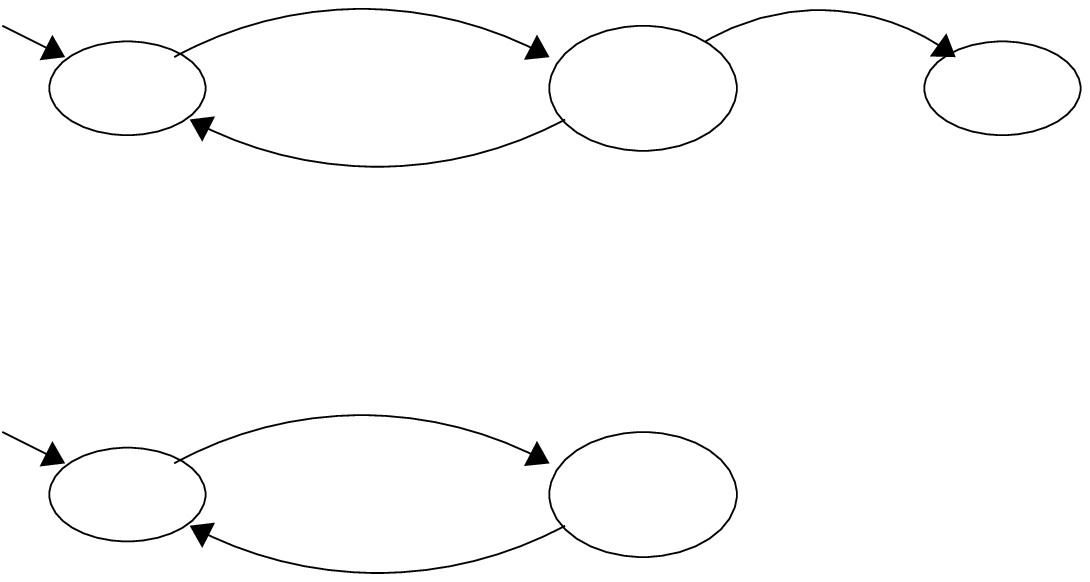}%
\end{picture}%
\setlength{\unitlength}{3947sp}%
\begingroup\makeatletter\ifx\SetFigFontNFSS\undefined%
\gdef\SetFigFontNFSS#1#2#3#4#5{%
  \reset@font\fontsize{#1}{#2pt}%
  \fontfamily{#3}\fontseries{#4}\fontshape{#5}%
  \selectfont}%
\fi\endgroup%
\begin{picture}(5195,3861)(664,-3289)
\put(976,-361){\makebox(0,0)[lb]{\smash{{\SetFigFontNFSS{12}{14.4}{\rmdefault}{\mddefault}{\updefault}{\color[rgb]{0,0,0}$\mbox{\tt idle}_1$}%
}}}}
\put(1501, 89){\makebox(0,0)[lb]{\smash{{\SetFigFontNFSS{12}{14.4}{\rmdefault}{\mddefault}{\updefault}{\color[rgb]{0,0,0}$x_1:=0;$}%
}}}}
\put(1501,239){\makebox(0,0)[lb]{\smash{{\SetFigFontNFSS{12}{14.4}{\rmdefault}{\mddefault}{\updefault}{\color[rgb]{0,0,0}$x_1 > 5$}%
}}}}
\put(1501,389){\makebox(0,0)[lb]{\smash{{\SetFigFontNFSS{12}{14.4}{\rmdefault}{\mddefault}{\updefault}{\color[rgb]{0,0,0}$!\mbox{\tt request}$}%
}}}}
\put(2926,-811){\makebox(0,0)[lb]{\smash{{\SetFigFontNFSS{12}{14.4}{\rmdefault}{\mddefault}{\updefault}{\color[rgb]{0,0,0}$?\mbox{\tt serve}$}%
}}}}
\put(5176,-361){\makebox(0,0)[lb]{\smash{{\SetFigFontNFSS{12}{14.4}{\rmdefault}{\mddefault}{\updefault}{\color[rgb]{0,0,0}$\mbox{\tt stop}_1$}%
}}}}
\put(4726,164){\makebox(0,0)[lb]{\smash{{\SetFigFontNFSS{12}{14.4}{\rmdefault}{\mddefault}{\updefault}{\color[rgb]{0,0,0}$x_1 > 10$}%
}}}}
\put(4726,314){\makebox(0,0)[lb]{\smash{{\SetFigFontNFSS{12}{14.4}{\rmdefault}{\mddefault}{\updefault}{\color[rgb]{0,0,0}$!\mbox{\tt end}$}%
}}}}
\put(2926,-961){\makebox(0,0)[lb]{\smash{{\SetFigFontNFSS{12}{14.4}{\rmdefault}{\mddefault}{\updefault}{\color[rgb]{0,0,0}$x_1:=0;$}%
}}}}
\put(901,-1261){\makebox(0,0)[lb]{\smash{{\SetFigFontNFSS{12}{14.4}{\rmdefault}{\mddefault}{\updefault}{\color[rgb]{0,0,0}(a) a model process $\calmdl$}%
}}}}
\put(2926,-2761){\makebox(0,0)[lb]{\smash{{\SetFigFontNFSS{12}{14.4}{\rmdefault}{\mddefault}{\updefault}{\color[rgb]{0,0,0}$?\mbox{\tt serve}$}%
}}}}
\put(1501,-1861){\makebox(0,0)[lb]{\smash{{\SetFigFontNFSS{12}{14.4}{\rmdefault}{\mddefault}{\updefault}{\color[rgb]{0,0,0}$x_2 > 5$}%
}}}}
\put(1501,-1711){\makebox(0,0)[lb]{\smash{{\SetFigFontNFSS{12}{14.4}{\rmdefault}{\mddefault}{\updefault}{\color[rgb]{0,0,0}$!\mbox{\tt request}$}%
}}}}
\put(2926,-2911){\makebox(0,0)[lb]{\smash{{\SetFigFontNFSS{12}{14.4}{\rmdefault}{\mddefault}{\updefault}{\color[rgb]{0,0,0}$x_2:=0;$}%
}}}}
\put(901,-3211){\makebox(0,0)[lb]{\smash{{\SetFigFontNFSS{12}{14.4}{\rmdefault}{\mddefault}{\updefault}{\color[rgb]{0,0,0}(b) a specification process $\calspc$}%
}}}}
\put(976,-2311){\makebox(0,0)[lb]{\smash{{\SetFigFontNFSS{12}{14.4}{\rmdefault}{\mddefault}{\updefault}{\color[rgb]{0,0,0}$\mbox{\tt idle}_2$}%
}}}}
\put(3451,-436){\makebox(0,0)[lb]{\smash{{\SetFigFontNFSS{12}{14.4}{\rmdefault}{\mddefault}{\updefault}{\color[rgb]{0,0,0}$x_1<20$}%
}}}}
\put(3376,-2161){\makebox(0,0)[lb]{\smash{{\SetFigFontNFSS{12}{14.4}{\rmdefault}{\mddefault}{\updefault}{\color[rgb]{0,0,0}$\mbox{\tt wait}_2$}%
}}}}
\put(3451,-2386){\makebox(0,0)[lb]{\smash{{\SetFigFontNFSS{12}{14.4}{\rmdefault}{\mddefault}{\updefault}{\color[rgb]{0,0,0}$x_2<15$}%
}}}}
\put(3376,-211){\makebox(0,0)[lb]{\smash{{\SetFigFontNFSS{12}{14.4}{\rmdefault}{\mddefault}{\updefault}{\color[rgb]{0,0,0}$\mbox{\tt wait}_1$}%
}}}}
\end{picture}%
\end{center} 
\caption{A model process and a specification process} 
\label{fig.ms_ne} 
\end{figure*}

{\example$\,$ \label{exmp.intro}}
In figure~\ref{fig.ms_ne}, 
we have the state-transition diagrams of two {\em timed automatas} ({\em TA})
\cite{AD94}.  
The one in figure~\ref{fig.ms_ne}(a) is for a model $\calmdl$ 
while the one in figure~\ref{fig.ms_ne}(b) is for a specification $\calspc$.  
We use ovals for the {\em control locations} of the TAs 
while arcs for the transition rules. 
In each oval, we label the invariance condition that must be satisfied 
in the location.  
For example, in location ${\tt wait}_1$, $\calmdl$ can stay for 
at most 20 time units. 
By each transition rule, we stack its synchronization event, 
triggering condition (guard), and actions.  
For convenience, tautology triggering conditions 
and nil actions are omitted. 
An event starting with a `?' represents a {\em receiving event} 
while one with a `!' represents a {\em sending event}.  
For example, for the transition from location $\mbox{\tt idle}_1$ to 
$\mbox{\tt wait}_1$, 
$\calmdl$ must send out an event request, 
be in a state satisfying $x_1>5$, and 
reset clock $x_1$ to zero. 
The specification in figure~\ref{fig.ms_ne}(b) does not simulate 
the model in figure~\ref{fig.ms_ne}(a) since event {\tt !end} of $\calmdl$ cannot 
be matched by any event of $\calspc$.  
Moreover $\calspc$ can neither receive a {\tt ?serve} event 
15 time units after issuing a {\tt !request} event 
while $\calmdl$ can.  
\qed 

However, the concept of simulation described in the last paragraph 
can be too restrictive in practice.  
Developers of a project usually cannot make too much assumption on the environment.  
The deadline constraints $x_1<20$ and $x_2<15$ 
can be too restrictive and hurt the extensibility 
of the model in the future.  
Another approach in this regard is using {\em fairness assumptions} 
\cite{GPVW95,SB00}.  
For example, for the model and specification processes in 
figure~\ref{fig.ms_ne}, 
we may want to check whether $\calspc$ simulates $\calmdl$ 
under the fairness assumption that the environment functions reasonably.  
Such an assumption can be captured with the fairness assumption 
that {\em there will always be infinitely many occurrences of 
event {\tt serve}}.  
Under this assumption, the $\calspc$ in figure~\ref{fig.ms_ne}(b) 
actually simulates the $\calmdl$ in figure~\ref{fig.ms_ne}(a). 

In this work, we propose the 
{\em simulation} with fairness assumptions for 
the processes in a dense-time setting.  
In such a setting, the model and the specification 
are both {\em generalized B\"uch timed automatas} 
({\em GBTA}) \cite{AD94} with 
communication channels and dense-time behaviors.  
We want to check whether the specification GBTA can 
simulate the model GBTA with multiple fairness assumptions.  
Following the approach of \cite{EL87,Wang04a}, 
we allow for the requirement and analysis 
of both strong and weak fairness assumptions.  
A {\em strong fairness} assumption intuitively 
means something will happen infinite many times.
A {\em weak fairness} assumption means something 
will hold true eventually forever.    
For convenience, we use two consecutive sets of formulas 
for fairness assumptions, 
the former for the strong fairness assumptions while 
the latter for the weak fairness assumptions.  

{\example$\,$\label{exmp.intro.fstate}} 
For the system in figure~\ref{fig.ms_ne}, 
we may have the following fairness assumptions.
\begin{center} 
$\{\mbox{\tt wait}_1\}\{\mbox{\tt idle}_1\vee\mbox{\tt wait}_1\}$ 
\end{center} 
The fairness assumptions in the above say that 
a valid computation of the system must satisfy  
the following two conditions. 
\begin{list1} 
\item {\em For the strong fairness assumption of} $\{\mbox{\tt wait}_1\}$: 
	For every $t\in \rnneg$, there exists a 
	$t'\in \rnneg$ with $t'>t$ such that 
	in the computation at time $t'$, 
	the model process is in location $\mbox{\tt wait}_1$. 
	This in fact says that the model must enter 
	location $\mbox{\tt wait}_1$ 
	infinitely many times along any valid computation.   
\item {\em For the weak fairness assumption of} 
	$\{\mbox{\tt idle}_1\vee\mbox{\tt wait}_1\}$:
	There exists a $t\in \rnneg$ such that for every  
	$t'\in \rnneg$ with $t'>t$, 
	the model process is in either locations $\mbox{\tt idle}_1$ or 
	$\mbox{\tt wait}_1$. 
	This in fact says that the model will stabilize 
	in locations $\mbox{\tt idle}_1$ and $\mbox{\tt wait}_1$.   
\end{list1} 
The two types of fairness assumption complement with each other 
and could be handy in making reasonable assumptions.  
\qed 

Furthermore, we also allow for both state formulas and 
event formulas \cite{Wang04a} 
in the description of fairness assumptions. 
State formulas are Boolean combinations of atomic statements 
of location names and state variables.  
For convenience, we use index $1$ for the model and 
index $2$ for the specification. 
Event formulas are then constructed with a precondition, 
a event name with a process index, and a post-condition in sequence.  

{\example$\,$\label{exmp.intro.fevent}} 
For the system in figure~\ref{fig.ms_ne}, 
we may write the following strong event fairness assumption. 
\begin{center} 
$\{(\mbox{\tt wait}_1)?\mbox{\tt serve}@(1)(\true)\}\{\}$  
\end{center} 
The event specification of $?\mbox{\tt serve}@(1)$  
means there is an event {\tt serve} received by process 1.  
The precondition for the event is $\mbox{\tt wait}_1$ while 
the post-condition is $\true$.  
The strong fairness assumption says that 
there should be infinite many events {\tt serve} 
received by process 1 in location $\mbox{\tt wait}_1$.  
\qed 

In general, 
an event specification can be either a receiving or a sending event.  
Such event formulas can be useful in making succinct specifications.  
Without such event formulas, we may have to use auxiliary state variables 
to distinguish those states immediately before (or after) an event 
from others.  
Such auxiliary variables usually unnecessarily exacerbate the state 
space explosion problem. 

One goal of our work is to develop a simulation-checking 
algorithm based on symbolic model-checking technology for 
dense-time systems \cite{HNSY92,Wang03}.  
To achieve this, we focus on a special class of simulations  
with the restriction of at most one fairness assumption  
for the specification.  
For convenience, we call this class 
the {\em USF} ({\em unit-specification-fairness}) 
{\em simulations}.  
Then we propose a symbolic algorithm for this special 
class of simulations.  
To our knowledge, this is the first such algorithm for GBTAs.  
Also unlike the fair simulation \cite{HKR97} checking algorithm based on 
ranking function in the literature, 
our algorithm is based on symbolic logic formulas manipulation, 
which has been proven useful in symbolic model checking \cite{BCMDH90}.  
Thus, our algorithm style can be interesting in itself.  

We also present a technique for the efficient simulation checking 
of concurrent systems by taking advantage of the 
common environment of a model and a specification. 
To apply the simulation checking algorithms mentioned in 
the above and in the literature \cite{Cerans92,TAKB96}, 
we need first construct a product automata of 
the environment $\calenv$ and the model $\calmdl$, 
in symbols $\calenv\times\calmdl$.  
Then we construct a product of $\calenv$ and the specification $\calspc$, 
in symbols $\calenv\times\calspc$.  
Then we check if $\calenv\times\calspc$ simulates 
$\calenv\times\calmdl$.   
As a result, such algorithms incur 
duplicate recording of the 
state information of $\calenv$ 
while manipulating representations for 
the simulation of $\calenv\times\calmdl$ by 
$\calenv\times\calspc$.  
Moreover, different transitions in $\calenv$ with the same 
observable events can also be matched in the simulation-checking. 
Such matching is not only counter-intuitive in simulation against 
the same environment, but also incur explosion in the enumeration 
of matched transitions between $\calenv\times\calmdl$ and 
$\calenv\times\calspc$.  
Our technique is embodied with the definition 
of a new simulation relation against a common environment.  
We have implemented this technique and experimented with 
benchmarks with and without fairness assumptions.  

We have the following presentation plan.  
Section~\ref{sec.relwork} is for related work. 
Section~\ref{sec.prel} reviews our system models \cite{AD94,Shaw92}. 
Sections~\ref{sec.simf} presents our simulation   
for dense-time systems with fairness assumptions.  
Section~\ref{sec.usf.neg.char} presents a characterization of 
the simulation when the specification is a B\"uchi TA.  
Section~\ref{sec.simf.alg} presents our simulation checking algorithm 
based on the characterization derived 
in section~\ref{sec.usf.neg.char}.  
Section~\ref{sec.sim.env} presents the simulation against 
a common environment and techniques for performance verification in this 
context.  
Sections~\ref{sec.imp} and \ref{sec.exp} 
respectively report our implementation and experiment. 
Section~\ref{sec.conc} is the conclusion.

\section{Related work} \label{sec.relwork}

Cerans showed that the bisimulation-checking problem 
of timed processes is decidable \cite{Cerans92}.  
Ta\c{s}{\scriptsize I}ran et al showed 
that the simulation-checking problem of dense-time automatas (TAs) 
\cite{AD94} is in EXPTIME \cite{TAKB96}.  
Weise and Lenzkes reported an algorithm based on zones for timed 
bisimulation checking \cite{WL97}. 
Cassez et al presented an algorithm for the reachability games of TAs 
with controllable and uncontrollable actions \cite{CDFL05}.  

Henzinger et al presented an algorithm that computes 
the time-abstract simulation that does not preserve timed 
properties \cite{HHK95}.  
Nakata also discussed how to do symbolic bisimulation checking 
with integer-time labeled transition systems \cite{Nakata97}. 
Beyer has implemented a refinement-checking algorithm for TAs 
with integer-time semantics \cite{Beyer00}.  

Lin and Wang presented a sound proof system 
for the bisimulation equivalence of TAs with dense-time semantics 
\cite{LW02}. 
Aceto et al discussed how to construct such a modal logic formula 
that completely characterizes a TA \cite{AIPP00}.

Larsen presented a similar theoretical framework for bisimulation 
in an environment for untimed systems \cite{Larsen85}.  
However no implementation that takes advantage of the common environment 
information for verification performance has been reported.  

Proposals for extending simulation with  
fair states have been discussed in \cite{GL94,HKR97,LT87}.  
Our simulation game of GBTAs 
stems from Henzinger et al's framework of fair simulation \cite{HKR97}.  
Techniques for simulation checking of GBAs 
were also discussed in \cite{GPVW95,SB00}.

\section{Preliminary} \label{sec.prel}

We have the following notations. 
$\rnall$ is the set of real numbers.
$\rnneg$ is the set of non-negative reals. 
$\nnneg$ is the set of nonnegative integers. 
Also `iff' is ``if and only if." 
Given a set $P$ of atomic propositions and a set $X$ of clocks, 
we use $\bbbbb(P,X)$ as the set of all Boolean combinations 
of logic atoms of the forms $q$ and $x\sim c$, where $q\in P$, $x\in X$, 
`$\sim$'$\in\{\leq, <,=,>,\geq\}$, and $c\in\nnneg$. 
An element in $\bbbbb(P,X)$ is called a {\em state-predicate}.

\subsection{Timed automata \label{subsec.ta}} 

A TA\cite{AD94,Shaw92,WHY06} is structured as a directed graph whose nodes 
are {\em modes (control locations)} and whose arcs are {\em transitions}. 
Please see figure~\ref{fig.ms_ne} for examples.  
A TA must always satisfy its {\em invariance condition}.   
Each transition is labeled with events, 
a {\em triggering condition}, and 
a set of clocks to be reset during the transitions. 
At any moment, a TA can stay in only one {\em mode}. 
If a TA executes a transition, 
then the triggering condition must be satisfied. 
In between transitions, all clocks in a TA increase 
their readings at a uniform rate.

{\definition \underline{\bf Timed automata (TA)}} 
A TA $A$ is a tuple  
$\langle Q, P, X, I, \lambda, E,\Sigma, \epsilon,\tau,\pi\rangle$. 
$Q$ is a finite set of modes (locations). 
$P$ is a finite set of propositions. 
$X$ is a finite set of clocks. 
$I\in \bbbbb(P,X)$ is the initial condition. 
$\lambda:Q\mapsto \bbbbb(P,X)$ is the invariance condition for each 
mode.
$E\subseteq Q\times Q$ is the set of process transitions.  
$\Sigma$ is a finite set of events. 
$\epsilon:E\mapsto 2^\Sigma$ is a mapping that defines the events 
at each transition.  
$\tau:E\mapsto \bbbbb(P,X)$ and $\pi:E\mapsto 2^X$ respectively 
define the triggering condition and the clock set to reset 
of each transition.  

Without loss of generality, 
we assume that for all $q,q'\neq Q$ with $q\neq q'$, 
$\lambda(q)\wedge\lambda(q')$ is a contradiction.  
We also assume that there is a null transition $\perp$  
that does nothing at any location.  
That is, the null transition transits from a location to the location itself. 
Moreover, $\tau(\perp)=\true$,  
$\pi(\perp)=\emptyset$, and $\epsilon(\perp)=\emptyset$. 
\qed

Given a TA $A=\langle Q,P,X,I,\lambda,E,\Sigma,\epsilon,\tau,\pi\rangle$, 
for convenience, we let 
$Q_A=Q$, $P_A=P$, $X_A=X$, $I_A=I$, $\lambda_A=\lambda$, $E_A=E$, 
$\Sigma_A=\Sigma$,  
$\epsilon_A = \epsilon$, $\tau_A = \tau$, and $\pi_A = \pi$. 
Also, for convenience, 
we let $V_A\defn \bigvee_{q\in Q_A}(\lambda_A(q))$ be 
the {\em invariance predicate} of $A$.  

{\example$\,$ \label{exmp.tas}}
We have already seen examples of TAs in figure~\ref{fig.ms_ne}.  
For the TA in figure~\ref{fig.ms_ne}(a), 
the attributes are listed in table~\ref{tab.ms_ne.a.attr}.  
\begin{table*}[t]
\begin{center} 
$\begin{array}{rcl} 
Q_\calmdl = P_\calmdl & = 
& \{\mbox{\tt idle}_1, \mbox{\tt wait}_1,\mbox{\tt stop}_1\} \\ 
X_\calmdl & = 
& \{x_1\} \\ 
I_\calmdl & \equiv
& \mbox{\tt idle}_1\wedge x_1=0 \\
\lambda_\calmdl & = & 
[\mbox{\tt idle}_1\mapsto \true, 
\mbox{\tt wait}_1\mapsto x_1<20, \mbox{\tt stop}_1\mapsto \true] \\ 
E_\calmdl & = & 
\{(\mbox{\tt idle}_1,\mbox{\tt wait}_1), 
(\mbox{\tt wait}_1,\mbox{\tt idle}_1),
(\mbox{\tt wait}_1,\mbox{\tt stop}_1)\} \\ 
\Sigma_\calmdl & = & 
\{\mbox{\tt request},\mbox{\tt serve},\mbox{\tt end}\} \\ 
\epsilon_\calmdl & = & 
[(\mbox{\tt idle}_1,\mbox{\tt wait}_1)\mapsto\{!\mbox{\tt request}\}, 
(\mbox{\tt wait}_1,\mbox{\tt idle}_1)\mapsto\{?\mbox{\tt serve}\}, 
(\mbox{\tt wait}_1,\mbox{\tt stop}_1)\mapsto\{!\mbox{\tt end}\}] \\ 
\tau_\calmdl & = & 
[(\mbox{\tt idle}_1,\mbox{\tt wait}_1)\mapsto x_1>5, 
(\mbox{\tt wait}_1,\mbox{\tt idle}_1)\mapsto \true, 
(\mbox{\tt wait}_1,\mbox{\tt stop}_1)\mapsto x_1>10] \\ 
\pi_\calmdl & = & 
[(\mbox{\tt idle}_1,\mbox{\tt wait}_1)\mapsto \{x_1\}, 
(\mbox{\tt wait}_1,\mbox{\tt idle}_1)\mapsto \{x_1\}, 
(\mbox{\tt wait}_1,\mbox{\tt stop}_1)\mapsto \emptyset] \\ 
\end{array}$\\[2mm] 
$f=[a_0\mapsto b_0, \ldots, a_k\mapsto b_k,\ldots]$ 
denotes a (partial or total) function $f$ 
with $f(a_0)=b_0, \ldots, f(a_k)=b_k, \ldots$.  
\end{center} 
\caption{Attributes of the TAs in figure~\ref{fig.ms_ne}(a)}
\label{tab.ms_ne.a.attr}
\end{table*} 
\qed 

A {\em valuation} of a set is a mapping from the set to another set. 
Given an $\eta\in \bbbbb(P, X)$ and a valuation $\nu$ of $X\cup P$, 
we say $\nu$ {\em satisfies} $\eta$, in symbols $\nu\models\eta$, 
iff $\eta$ is evaluated $\true$ when the variables in $\eta$ 
are interpreted according to $\nu$.  

{\definition \underline{\bf States of a TA}} 
Suppose we are given a TA $A$. 
A {\em state} $\nu$ of $A$ is a valuation of $X_A\cup P_A$ 
with the following constraints. 
\begin{list1}
\item For each $p\in P_A$, 
	$\nu(p)\in \{\false,\true\}$. 
	There exists a $q\in Q_A$ such that 
	$\nu\models\lambda(q)$ and 
	for all $q'\neq q$, $\nu\not\models\lambda(q')$.   
	Given a $q\in Q_A$, if $\nu\models \lambda(q)$, 
	we denote $q$ as $\ttmode_A(\nu)$.  
\item For each $x\in X_A$, $\nu(x)\in \rnneg$.
\end{list1}
In addition, we require that $\nu\models V_A$.  
We let $\bbbbs\langle S\rangle$ denote the set of states of $A$. 
\qed

Note that we define a state as a mapping instead 
of as a pair of control locations and a real mapping as in \cite{ACD90}.  
This is for the convenience of presentation when latter 
we want to discuss the state-pairs in simulation relations.  

For any state $\nu$ and real number $t\in \rnneg$, 
$\nu+t$ is a state identical to $\nu$ 
except that for every clock $x\in X_A$, 
$(\nu+t)(x) = \nu(x)+t$.  
Also given a process transition $e=(q,q')\in E_A$, 
we use $\nu e$ to denote the destination state from $\nu$ 
through the execution of $e$.  
Formally, if $\nu\models \tau_A(e)$, 
then $\nu e$ 
is a new state that is identical to $\nu$ 
except that the following constraints are true. 
\begin{list1}
\item $q=\ttmode_A(\nu)$ and $q'=\ttmode_A(\nu e)$.  
\item For every clock $x\in\pi_A(e)$, $\nu e(x)=0$. 
\item For every clock $x\not\in\pi_A(e)$, $\nu e(x)=\nu(x)$. 
\end{list1}

Given a $t\in\rnneg$ and a transition $e$, 
we write $\nu \stackrel{t,e}{\longrightarrow} \nu'$ 
iff $\nu+t\models \tau_A(e)$, $(\nu+t)e=\nu'$, 
$\nu'\models V_A$, and for each $t'\in [0,t]$, 
$\nu+t'\models V_A$. 
For convenience, we use $[\nu \stackrel{t,e}{\longrightarrow}]$ 
to denote such a $\nu'$ with $\nu \stackrel{t,e}{\longrightarrow} \nu'$.

{\definition \underline{\bf Runs}} 
A {\em run} of a TA 
$A$ is an infinite sequence of state-transition-time triples 
$(\nu_0,e_0,t_0)(\nu_1,e_1,t_1)\ldots(\nu_k,e_k,t_k)
\ldots\ldots$ 
with the following restrictions.  
\begin{list1} 
\item {\bf Non-Zeno requirement:} 
	$t_0 t_1 \ldots t_k\ldots\ldots$ is a non-decreasing and divergent 	real-number sequence. 
  	That is, 
  	$\forall k\in\nnneg, t_k\leq t_{k+1}$ and 
  	$\forall c\in \nnneg, \exists k>1, t_k>c$. 
\item 
	For all $k\in\nnneg$, either 
	$\nu_k+t_{k+1}-t_k = \nu_{k+1}$ or 
	$\nu_k\stackrel{t_{k+1}-t_k,e_{k+1}}{\lllrarrow} \nu_{k+1}$.  
\end{list1}  
A {\em run prefix} is a finite prefix of a run.  
A run prefix or a run $(\nu_0,e_0,t_0)\ldots$ of $A$ is {\em initial} 
iff $\nu_0\models I_A$.  
\qed

\subsection{Generalized B\"uchi TAs} 

Suppose we are given a TA $\cala$.  
An {\em event-predicate} is of the form     
$\eta_1 a\eta_2$.   
Here   
$\eta_1$ and $\eta_2$ are two state-predicates in 
$\bbbbb(P_\cala,X_\cala)$ respectively for 
the precondition and the post-condition of the event.  
$a\in \Sigma_\cala$ is an event name.  
Event-predicate ``$\eta_1 a\eta_2$" specifies the observation of 
event $a$ with precondition $\eta_1$ 
and post-condition $\eta_2$.   

In this work, we allow  
{\em fairness assumptions} either as state-predicates or as 
event-predicates. 
A state fairness assumption is in $\bbbbb(P_\cala,X_\cala)$.  
An event fairness assumption is an event-predicate of $\cala$.  
Given two sets $\Phi$ and $\Psi$ of fairness assumptions, 
$\Phi\Psi$ denotes a {\em multi-fairness assumption} 
({\em MF-assumption}) 
for $\cala$.   
All elements in $\Phi$ are called {\em strong fairness assumptions} 
while all in $\Psi$ are called {\em weak fairness assumptions}.  
A run $(\nu_0,e_0,t_0)\ldots(\nu_k,e_k,t_k)\ldots$ 
of $\cala$ satisfies $\Phi\Psi$ 
iff the following constraints hold. 
\begin{list1} 
\item For every state-predicate $\eta\in \Phi$, 
	there are infinitely many $k$'s such that 
	for some $t\in [0,t_{k+1}-t_k]$, $\nu_k+t\models\eta$.  
\item For every event-predicate $\eta_1 a\eta_2$ in $\Phi$, 
	there are infinitely many $k$'s such that   
	$\nu_h+(t_{h+1}-t_h)\models\eta_1$,  
	$a\in\epsilon_\cala(e_{h+1})$, and 
	$\nu_{h+1}\models \eta_2$.  
\item For every state-predicate $\eta\in \Psi$, 
	there is a $k$ such that for every
	$h>k$ and $t\in [0,t_{h+1}-t_h]$, 
	$\nu_h+t\models\eta$.  
\item For every event-predicate $\eta_1 a\eta_2$ in $\Psi$, 
	there is a $k$ such that for every
	$h>k$,  
	if $\nu_h+(t_{h+1}-t_h)\models\eta_1$ and 
	$a\in\epsilon_\cala(e_{h+1})$, 
	then $\nu_{h+1}\models \eta_2$.  
\end{list1} 
Given a TA $\cala$ and a state $\nu\in \bbbbs\langle \cala\rangle$, 
we let $\Omega_\cala(\nu,\Phi\Psi)$ 
denote the set of runs of $\cala$ from $\nu$ 
satisfying $\Phi\Psi$.  
The following definition shows how to formally model 
real-time systems with fairness assumptions.

{\definition \underline{\bf GBTAs and BTAs}
\label{def.gbcta}} 
A {\em generalized B\"uchi TA} ({\em GBTA})
is a pair $\langle \cala,\Phi\Psi\rangle$ with 
a TA $\cala$ and an MF-assumption $\Phi\Psi$. 
If $|\Phi|+|\Psi|\leq 1$, the pair is also called a {\em B\"uchi TA} 
({\em BTA}).  
\qed

{\example$\,$ \label{exmp.ctas}}
For the model $\calmdl$ in figure~\ref{fig.ms_ne}(a), 
we may have a GBTA 
\begin{center} 
$\langle \calmdl,
  \{\mbox{\tt wait}_1,\true?\mbox{\tt serve}@(2)\mbox{\tt idle}_1\}
  \emptyset\rangle$ 
\end{center} 
that assumes $\calmdl$ should stay in 
location $\mbox{\tt wait}_1$ infinitely many times 
and event $\mbox{\tt serve}$ should be received by 
$\calmdl$ infinitely many times with post-condition $\mbox{\tt idle}_1$.  

We may also have the following GBTA
\begin{center} 
$\langle \calmdl,
  \emptyset\{\mbox{\tt stop}_1\}\rangle$ 
\end{center} 
that assumes that $\calmdl$ should eventually stabilize in location 
$\mbox{\tt stop}_1$.  
\qed

\section{Simulation of GBTAs 
\label{sec.simf} 
}

Suppose we are given two TAs $\cala,\calb$.  
For any transitions $e\in E_\cala$ and $f\in E_\calb$, 
$e$ and $f$ are {\em compatible} 
iff $\epsilon_\cala(e)=\epsilon_\calb(f)\neq \emptyset$. 
That is, the observable events of 
the two automatas on the two transitions must be nontrivially identical. 
For each $e\in E_\cala$ with $\epsilon_\cala(e)\neq\emptyset$, 
we use $E^{(e)}_\calb$ to denote the subset 
of $E_\calb$ with elements compatible with $e$.  
For each $e\in E_\cala-\{\perp\}$ with $\epsilon_\cala(e)=\emptyset$, 
$E^{(e)}_\calb=\{\perp\}$.  
Also, $E^{(\perp)}_\calb$ denotes the subset of $E_\calb$ 
with elements $f$ such that $\epsilon_\calb(f)=\emptyset$. 

In this section, from now on, 
we assume the context of two GBTAs 
$\langle\calmdl,\Phi_\calmdl\Psi_\calmdl\rangle$ and 
$\langle\calspc,\Phi_\calspc\Psi_\calspc\rangle$ 
respectively for the model and the specification.  

Given a state $\mu$ of $\calmdl$ and 
a state $\nu$ of $\calspc$,    
we use $\mu\nu$ to denote the state-pair of $\mu$ and $\nu$.   
Operationally, $\mu\nu$ can be viewed as $\mu\circ\nu$, 
the functional composition of $\mu$ and $\nu$.  
A {\em play} between $\calmdl$ and $\calspc$  
is made of two matching runs, 
one of $\calmdl$ and the other of $\calspc$.  
Conceptually, it is a sequence 
\begin{center} 
$(\mu_0\nu_0,e_0f_0,t_0)\ldots(\mu_k\nu_k,e_kf_k,t_k)\ldots$ 
\end{center} 
of triples with the following restrictions.  
\begin{list1} 
\item $(\mu_0,e_0,t_0)\ldots(\mu_k,e_k,t_k)\ldots$ 
	is a run of $\calmdl$.   
	For convenience, we denote this run as  
	$\mbox{\em run}_\calmdl(\rho)$. 
\item $(\nu_0,f_0,t_0)\ldots(\nu_k,f_k,t_k)\ldots$ 
	is a run of $\calspc$.  
	For convenience, we denote this run as  
	$\mbox{\em run}_\calspc(\rho)$. 
\item For each $k\in \nnneg$, $f_k\in E^{(e_k)}_\calspc$. 
\end{list1} 
The play is {\em initial} iff 
$\mu_0\models I_\calmdl$ and $\nu_0\models I_\calspc$.  
A {\em play prefix} is a finite prefix of a play.  
Given a play $\rho$, 
we let $\rho^{(k)}$ be the prefix represented as 
the sequence of the first $k+1$ elements of $\rho$.   

Given a run (prefix) 
\begin{center} 
$\theta=(\mu_0,e_0,t_0)\ldots(\mu_k,e_k,t_k)\ldots$ 
\end{center} 
of 
$\calmdl$ and 
a play (prefix) 
\begin{center} 
$\rho=(\bar\mu_0\bar\nu_0,\bar{e}_0\bar{f}_0,\bar{t}_0)
\ldots(\bar\mu_h\bar\nu_h,\bar{e}_h\bar{f}_h,\bar{t}_h)\ldots$ 
\end{center} 
between  
$\calmdl$ and $\calspc$, 
we say $\rho$ {\em embeds} $\theta$ iff 
there is a monotonically increasing integer function $\gamma()$ 
such that $\gamma(0)=0$ and 
for each $k\in\nnneg$, $\bar\mu_{\gamma(k)}=\mu_k$, 
	$\bar{e}_{\gamma(k)}=e_k$,  
	$\bar{t}_{\gamma(k)}=t_k$, and  
	for each $h\in (\gamma(k),\gamma(k+1))$, $\bar{e}_h=\perp$.    
Notationally, we let $\rho\rhd_\calmdl\theta$ denote 
the embedding relation between $\rho$ and $\theta$.  
Similarly we can define $\rho\rhd_\calspc\theta'$ 
for the embedding relation between $\rho$ and a run $\theta'$ of $\calspc$.  

A {\em strategy} in a game tells a TA what to execute at a state-pair 
in a play that is developing.  
Specifically, 
a {\em strategy} $\sigma$ for $\calspc$ is a mapping 
from play prefixes of $\calmdl$ and $\calspc$ to event sets 
of $\Sigma_\calspc$.  
Symmetrically, we can define strategies for $\calmdl$.  
Given a strategy $\sigma$ for $\calspc$ 
and a play 
$\rho=(\mu_0\nu_0,e_0f_0,t_0)\ldots(\mu_k\nu_k,e_kf_k,t_k)\ldots$ 
between $\calmdl$ and $\calspc$, 
we say that $\rho$ {\em complies} to $\sigma$ 
iff the following constraints are satisfied.  
\begin{list1} 
\item For each $k\in \nnneg$ and $t\in [0,t_{k+1}-t_k)$, 
	\begin{center} 
	$\sigma(\rho^{(k)}
		((\mu_k+t)(\nu_k+t),\perp\perp,t_k+t))=\perp$.  
	\end{center} 
\item For each $k\in \nnneg$ and $t=t_{k+1}-t_k$
	with either $t_{k+2}-t_{k+1}>0$ or 
	$f_{k+1}\neq \perp$, 
	\begin{center} 
	$\sigma(\rho^{(k)}((\mu_k+t)(\nu_k+t),
		\perp\perp,
		t_{k+1}))
		=f_{k+1}$.  
	\end{center} 
\end{list1} 
Similarly, we can also define the compliance of plays to strategies of 
$\calmdl$.  
Given a state-pair 
$\mu\nu\in \bbbbs\langle\calmdl\rangle\times \bbbbs\langle\calspc\rangle$, 
a run $\theta$ of $\calmdl$ from $\mu$, 
and a strategy $\sigma$ of $\calspc$, 
we let $\rho=\emplay(\mu\nu,\theta,\sigma)$ 
be the play (prefix) from $\mu\nu$ 
with the following restrictions. 
\begin{list1} 
\item $\rho$ complies to $\sigma$.
\item If $\rho$ is of infinite length, 
	then it embeds $\theta$.  
\item If $\rho$ is of finite length, 
	then there is a finite prefix 
	$\bar\theta=(\mu_0,e_0,t_0)\ldots(\mu_k,e_k,t_k)$ of $\theta$ 
	with the following restrictions.
	\begin{list2} 
	\item $\rho$ embeds $\bar\theta$. 
	\item Any prefix of $\theta$ that supersedes $\bar\theta$ is not 
		embedded by $\rho$.  
	\end{list2} 
\end{list1} 
Note that it may happen that $\emplay(\mu\nu,\theta,\sigma)$
is of only finite length.  
This can happen when at the end of the finite play, a player 
chooses a transition with an event set that the other player 
({\em opponent}) cannot choose 
a transition to match. 
This can also happen when at the end of the finite play, 
a player can only execute matching transitions with post-condition
falling outside the invariance predicate.

{\definition \label{def.simf} 
\underline{\bf Simulation of GBTAs}} 
A {\em simulation} $F$ of 
$\langle\calmdl,\Phi_\calmdl\Psi_\calmdl\rangle$ by 
$\langle\calspc,\Phi_\calspc\Psi_\calspc\rangle$   
is a binary relation 
$F\subseteq \bbbbs\langle\calmdl\rangle\times\bbbbs\langle\calspc\rangle$ 
such that for every $\mu\nu\in F$ and every 
run $\theta$ of $\calmdl$ from $\mu$ that satisfies 
$\Phi_\calmdl\Psi_\calmdl$,  
there exists a play $\rho$ from $\mu\nu$ 
such that $\rho$ embeds $\theta$ and 
$\mbox{\em run}_\calspc(\rho)$ satisfies $\Phi_\calspc\Psi_\calspc$.  

We say that 
$\langle\calspc,\Phi_\calspc\Psi_\calspc\rangle$ simulates 
$\langle\calmdl,\Phi_\calmdl\Psi_\calmdl\rangle$,  
in symbols $\langle\calmdl,\Phi_\calmdl\Psi_\calmdl\rangle
\propto \langle\calspc,\Phi_\calspc\Psi_\calspc\rangle$, 
if there exists a simulation $F$ of 
$\langle\calmdl,\Phi_\calmdl\Psi_\calmdl\rangle$ by 
$\langle\calspc,\Phi_\calspc\Psi_\calspc\rangle$ 
such that 
for every $\mu\models I_\calmdl\wedge V_\calmdl$, 
there exists a $\nu\models I_\calspc\wedge V_\calspc$ 
with $\mu\nu\in F$.  
\qed

{\example$\,$ \label{exmp.simf}}
For the TAs in figure~\ref{fig.ms_ne}, 
we have that 
$\langle \calspc,\emptyset\emptyset\rangle$ 
does not simulate 
$\langle \calmdl,\emptyset\emptyset\rangle$.  
Also, $\langle \calspc,\{\true?\mbox{\tt serve}\true\}\emptyset\rangle$ 
does not simulate 
$\langle \calmdl,\emptyset\{\mbox{\tt stop}_1\}\rangle$.  
However, 
$\langle \calspc,\emptyset\emptyset\rangle$ 
simulates 
$\langle \calmdl,\{\mbox{\tt wait}_1\}\emptyset\rangle$.  
\qed 

If $\langle\calspc,\Phi_\calspc\Psi_\calspc\rangle$  
simulates  
$\langle\calmdl,\Phi_\calmdl\Psi_\calmdl\rangle$, 
then 
for all initial states $\mu$ and runs $\theta$ of $\calmdl$ from $\mu$ 
satisfying $\Phi_\calmdl\Psi_\calmdl$, 
there exists a strategy $\sigma$ such that 
$\emplay(\mu\nu, \theta,\sigma)$ satisfies 
$\Phi_\calspc\Psi_\calspc$.  
We call such a $\sigma$ a {\em simulating strategy} for 
$\theta$ by $\calspc$.  

If $\langle\calspc,\Phi_\calspc\Psi_\calspc\rangle$ does not 
simulate  
$\langle\calmdl,\Phi_\calmdl\Psi_\calmdl\rangle$, 
then 
there exists an initial run $\theta$ of $\calmdl$ such that 
$\theta$ satisfies $\Phi_\calmdl\Psi_\calmdl$ and 
for all initial states $\nu$ and all strategies $\sigma$ of $\calspc$, 
all initial runs of $\calspc$ embedded by  
$\emplay(\mu\nu, \theta,\sigma)$ do not satisfy 
$\Phi_\calspc\Psi_\calspc$.  
We call such a run $\theta$ a {\em refuting run} of $\calmdl$.  

A strategy $\sigma$ of a TA $\calspc$ is {\em memory-less} iff 
for any two plays $\rho$ and $\rho'$ that end at the same triple, 
$\sigma(\rho)=\sigma(\rho')$.  
It is well known that parity games and reachability games all have 
memory-less winning strategies for either player \cite{GTW02}.  
The following lemma shows that the simulation  
of GBAs may need finite-memory refuting strategies.  

\begin{figure}[t] 
\begin{center}
\begin{picture}(0,0)%
\includegraphics{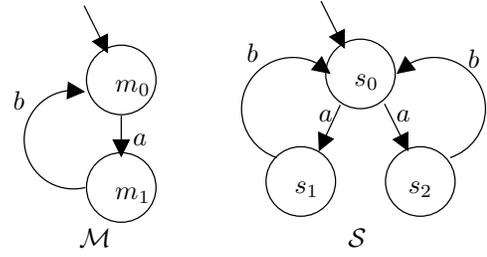}%
\end{picture}%
\setlength{\unitlength}{4144sp}%
\begingroup\makeatletter\ifx\SetFigFontNFSS\undefined%
\gdef\SetFigFontNFSS#1#2#3#4#5{%
  \reset@font\fontsize{#1}{#2pt}%
  \fontfamily{#3}\fontseries{#4}\fontshape{#5}%
  \selectfont}%
\fi\endgroup%
\begin{picture}(2847,1550)(166,-863)
\put(1863,-476){\makebox(0,0)[lb]{\smash{{\SetFigFontNFSS{10}{12.0}{\rmdefault}{\mddefault}{\updefault}{\color[rgb]{0,0,0}$s_1$}%
}}}}
\put(2543,-476){\makebox(0,0)[lb]{\smash{{\SetFigFontNFSS{10}{12.0}{\rmdefault}{\mddefault}{\updefault}{\color[rgb]{0,0,0}$s_2$}%
}}}}
\put(2471,-47){\makebox(0,0)[lb]{\smash{{\SetFigFontNFSS{10}{12.0}{\rmdefault}{\mddefault}{\updefault}{\color[rgb]{0,0,0}$a$}%
}}}}
\put(2006,-47){\makebox(0,0)[lb]{\smash{{\SetFigFontNFSS{10}{12.0}{\rmdefault}{\mddefault}{\updefault}{\color[rgb]{0,0,0}$a$}%
}}}}
\put(2901,275){\makebox(0,0)[lb]{\smash{{\SetFigFontNFSS{10}{12.0}{\rmdefault}{\mddefault}{\updefault}{\color[rgb]{0,0,0}$b$}%
}}}}
\put(1577,311){\makebox(0,0)[lb]{\smash{{\SetFigFontNFSS{10}{12.0}{\rmdefault}{\mddefault}{\updefault}{\color[rgb]{0,0,0}$b$}%
}}}}
\put(2221,167){\makebox(0,0)[lb]{\smash{{\SetFigFontNFSS{10}{12.0}{\rmdefault}{\mddefault}{\updefault}{\color[rgb]{0,0,0}$s_0$}%
}}}}
\put(2185,-798){\makebox(0,0)[lb]{\smash{{\SetFigFontNFSS{10}{12.0}{\rmdefault}{\mddefault}{\updefault}{\color[rgb]{0,0,0}$\calspc$}%
}}}}
\put(789,132){\makebox(0,0)[lb]{\smash{{\SetFigFontNFSS{10}{12.0}{\rmdefault}{\mddefault}{\updefault}{\color[rgb]{0,0,0}$m_0$}%
}}}}
\put(789,-512){\makebox(0,0)[lb]{\smash{{\SetFigFontNFSS{10}{12.0}{\rmdefault}{\mddefault}{\updefault}{\color[rgb]{0,0,0}$m_1$}%
}}}}
\put(897,-190){\makebox(0,0)[lb]{\smash{{\SetFigFontNFSS{10}{12.0}{\rmdefault}{\mddefault}{\updefault}{\color[rgb]{0,0,0}$a$}%
}}}}
\put(181, 24){\makebox(0,0)[lb]{\smash{{\SetFigFontNFSS{10}{12.0}{\rmdefault}{\mddefault}{\updefault}{\color[rgb]{0,0,0}$b$}%
}}}}
\put(575,-798){\makebox(0,0)[lb]{\smash{{\SetFigFontNFSS{10}{12.0}{\rmdefault}{\mddefault}{\updefault}{\color[rgb]{0,0,0}$\calmdl$}%
}}}}
\end{picture}%
\end{center}
\caption{A simulation game with winning strategies of $\calmdl$ that need memory.}
\label{fig.simf.memory}
\end{figure}

{\lemma \label{lemma.simf.memory} 
There is a simulation of GBAs
with a simulation strategy for the specification but 
without a memory-less simulation strategy for the specification. 
} 
\\\pf  
In figure~\ref{fig.simf.memory}, 
we have the TAs of two GBAs 
$\langle\calmdl,\{m_0,m_1\}\emptyset\rangle$ and 
$\langle\calspc,\{s_1,s_2\}\emptyset\rangle$.  
Suppose we have a state-pair $\mu\nu$ with 
$\ttmode_\calmdl(\mu)=m_0$ and 
$\ttmode_\calspc(\nu)=s_0$.  
As can be seen, for any memory-less strategy $\sigma$, 
either transition $(s_0, s_1)$ will always 
be chosen for 
any initial play prefix that ends at $\mu\nu$ 
or transition $(s_0, s_2)$ will always be.  
But such plays do not satisfy the strong fairness assumption of 
$\langle\calspc,\{s_1,s_2\}\emptyset\rangle$ and 
cannot be used to fulfill the strong fairness assumptions of $\calspc$.  
Thus we know there is no memory-less simulation strategy 
for $\langle\calspc,\{s_1,s_2\}\emptyset\rangle$.  

On the other hand, we can devise a strategy for $\calspc$ that chooses 
$(s_0, s_1)$ and $(s_0, s_2)$ alternately. 
It is clear that such a strategy fulfills   
the strong fairness assumptions of $\{s_1,s_2\}$.  
\qed

\section{Characterization of USF-simulation 
\label{sec.usf.neg.char}
} 

In this work, we focus on characterization of 
the simulation of a model GBTA by a specification BTA.  
That is, we restrict that the specification 
$\langle \calspc,\Phi_\calspc\Psi_\calspc\rangle$ 
is a BTA with $|\Phi_\calspc|+|\Psi_\calspc|\leq 1$.  

For convenience, given an MF-assumption $\Phi\Psi$ and 
a play $\rho=(\mu_0\nu_0,e_0f_0,t_0)\ldots(\mu_k\nu_k,e_kf_k,t_k)\ldots$, 
we may also define the satisfaction of $\Phi\Psi$ by $\rho$ 
in a way similar to the satisfaction of $\Phi\Psi$ by runs. 

According to definition~\ref{def.simf}, 
a state-pair $\mu\nu$ is not in any simulation if 
there exists a run $\theta$ of $\calmdl$ from $\mu$, satisfying 
$\Phi_\calmdl\Psi_\calmdl$, such that 
for every strategy $\sigma$ for $\calspc$ and 
play $\rho$ from $\mu\nu$ 
complying to $\sigma$ and embedding $\theta$, 
$\rho$ does not satisfy $\Phi_\calspc\Psi_\calspc$.  
Put this description in a structural way, we have the following
presentation. 
\begin{center}
$\begin{array}{ll}
	& (\mu\mbox{ starts a run $\theta$ of $\calmdl$ satisfying }\Phi_\calmdl\Psi_\calmdl)\\
\wedge	& \forall\rho\left(\begin{array}{l} 
	\rho\mbox{ starts from }\mu\nu\mbox{ and embeds }\theta. \\
	\hspace*{2mm}\Rightarrow
	 \rho\mbox{ does not satisfy }
		\Phi_\calspc\Psi_\calspc.
	\end{array}\right)
\end{array}
$
\end{center} 
According to the composition of $\Phi_\calspc\Psi_\calspc$, 
this can be broken down to cases described with the following four lemmas. 

{\lemma \label{lemma.rsim.ss} 
In case $\Phi_\calspc=\{\eta\}$ for a state-predicate $\eta$,  
a state-pair $\mu\nu$ is not in any simulation of 
$\langle \calmdl,\Phi_\calmdl\Psi_\calmdl\rangle$ 
by $\langle \calspc,\Phi_\calspc\Psi_\calspc\rangle$ iff   
\begin{center}
$\renewcommand{\arraycolsep}{0pt}
\begin{array}{ll}
	& (\mu\mbox{ starts a run $\theta$ of $\calmdl$ satisfying }\Phi_\calmdl\Psi_\calmdl)\\
\wedge	& \forall\rho\left(\begin{array}{l} 
	\rho\mbox{ starts from }\mu\nu\mbox{ and embeds }\theta. \\
	\hspace*{2mm}\Rightarrow
	\rho\mbox{ satisfies }
		\Phi_\calmdl(\Psi_\calmdl\cup\{\neg\eta\}).
	\end{array}\right)
\end{array}
$
\end{center} 
is true. 
}
\\\pf 
According to the argument in the beginning of the subsection, 
we only have to prove that the following two statements are equivalent 
in the context that $\rho$ embeds $\theta$. 
\begin{list1} 
\item $\rho\mbox{ does not satisfy }\{\eta\}\emptyset$.
\item $\rho\mbox{ satisfies }
		\Phi_\calmdl(\Psi_\calmdl\cup\{\neg\eta\})$.
\end{list1} 
Assume that 
\begin{center} 
$\rho=(\mu_0\nu_0,e_0f_0,t_0)\ldots(\mu_k\nu_k,e_kf_k,t_k)\ldots$.  
\end{center} 
We can prove this equivalence in two directions. 

\noindent 
$(\Rightarrow)$ 
We assume that $\rho\mbox{ does not satisfy }\{\eta\}\emptyset$.  
According to the definition of strong fairness, we know that 
there are only finitely many $k$'s with a $t\in [0,t_{k+1}-t_k]$ 
such that $\mu_k\nu_k+t\models \eta$.  
We let $m$ the maximum of such $k$'s. 
Then it is clear that for every $h>m$ and $t\in [0,t_{h+1}-t_h]$, 
$\mu_h\nu_h+t\not\models \eta$.  
This means that $\rho$ satisfies $\emptyset\{\neg\eta\}$.
Then the embedding of $\theta$ by $\rho$ 
implies that $\rho$ satisfies $\Phi_\calmdl(\Psi_\calmdl\cup\{\neg\eta\})$. 

\noindent 
$(\Leftarrow)$ 
We assume that 
$\rho\mbox{ satisfies }\Phi_\calmdl(\Psi_\calmdl\cup\{\neg\eta\})$.  
Then according to the definition of weak fairness, 
we know that 
there exists an $m$ such that for every $h>m$ and $t\in [0,t_{h+1}-t_h]$, 
$\mu_h\nu_h+t\models \neg\eta$.  
Thus it is not true that there are infinitely many $k$'s with 
a $t\in[0,t_{k+1}-t_k]$ such that $\mu_k\nu_k+t\models \eta$.   
According to the definition of strong fairness,  
$\rho\mbox{ does not satisfy }\{\eta\}\emptyset$.  

With the proof of the two directions, we know the lemma is proven. 
\qed 

{\lemma \label{lemma.rsim.se} 
In case $\Phi_\calspc=\{\eta_1 a\eta_2\}$ 
for an event-predicate $\eta_1 a\eta_2$,  
a state-pair $\mu\nu$ is not in any simulation of 
$\langle \calmdl,\Phi_\calmdl\Psi_\calmdl\rangle$ 
by $\langle \calspc,\Phi_\calspc\Psi_\calspc\rangle$ iff  
\begin{center}
$\renewcommand{\arraycolsep}{0pt}
\begin{array}{ll}
	& (\mu\mbox{ starts a run $\theta$ of $\calmdl$ satisfying }\Phi_\calmdl\Psi_\calmdl)\\
\wedge	& \forall\rho\left(\begin{array}{l} 
	\rho\mbox{ starts from }\mu\nu\mbox{ and embeds }\theta. \\
	\hspace*{2mm}\Rightarrow
	\rho\mbox{ satisfies }
		\Phi_\calmdl(\Psi_\calmdl\cup\{\eta_1 a\neg\eta_2\}).
	\end{array}\right)
\end{array}
$
\end{center} 
is true. 
}
\\\pf 
Suppose we are given 
\begin{center} 
$\rho=(\mu_0\nu_0,e_0f_0,t_0)\ldots(\mu_k\nu_k,e_kf_k,t_k)\ldots$.  
\end{center} 
The proof is similar to the one for lemma~\ref{lemma.rsim.ss} except that 
we need to show that for a $k\geq 0$, the equivalence between 
the following two statements. 
\begin{list1} 
\item It is not true that 
	$(\mu_k+t_{k+1}-t_k)(\nu_k+t_{k+1}-t_k)\models \eta_1$, 
	$a\in \epsilon_\calmdl\cap\epsilon_\calspc$, 
	and $\mu_{k+1}\nu_{k+1}\models \eta_2$.  
\item If $(\mu_k+t_{k+1}-t_k)(\nu_k+t_{k+1}-t_k)\models \eta_1$ and 
	$a\in \epsilon_\calmdl\cap\epsilon_\calspc$, then 
	$\mu_{k+1}\nu_{k+1}\not\models \eta_2$.  
\end{list1} 
This equivalence follows from the semantics of propositional logic.  
By treating the event-predicate as a state-predicate, we can 
prove the lemma as we have proved lemma~\ref{lemma.rsim.ss}.  
\qed

{\lemma \label{lemma.rsim.ws} 
In case $\Psi_\calspc=\{\eta\}$ for a state predicate $\eta$,  
a state-pair $\mu\nu$ is not in any simulation of 
$\langle \calmdl,\Phi_\calmdl\Psi_\calmdl\rangle$ 
by $\langle \calspc,\Phi_\calspc\Psi_\calspc\rangle$ iff  
\begin{center}
$\renewcommand{\arraycolsep}{0pt}
\begin{array}{ll}
	& (\mu\mbox{ starts a run $\theta$ of $\calmdl$ satisfying }\Phi_\calmdl\Psi_\calmdl)\\
\wedge	& \forall\rho\left(\begin{array}{l} 
	\rho\mbox{ starts from }\mu\nu\mbox{ and embeds }\theta. \\
	\hspace*{2mm}\Rightarrow
	\rho\mbox{ satisfies }
		(\Phi_\calmdl\cup\{\neg\eta\})\Psi_\calmdl.
	\end{array}\right)
\end{array}
$
\end{center} 
is true. 
}
\\\pf 
By replacing $\eta$ with $\neg\eta$, 
we can use a proof similar to the one for lemma~\ref{lemma.rsim.ss} 
for this lemma.  
\qed

{\lemma \label{lemma.rsim.we} 
In case $\Psi_\calspc=\{\eta_1 a\eta_2\}$ for a state predicate $\eta$,  
a state-pair $\mu\nu$ is not in any simulation of 
$\langle \calmdl,\Phi_\calmdl\Psi_\calmdl\rangle$ 
by $\langle \calspc,\Phi_\calspc\Psi_\calspc\rangle$ iff  
\begin{center}
$\renewcommand{\arraycolsep}{0pt}
\begin{array}{ll}
	& (\mu\mbox{ starts a run $\theta$ of $\calmdl$ satisfying }\Phi_\calmdl\Psi_\calmdl)\\
\wedge	& \forall\rho\left(\begin{array}{l} 
	\rho\mbox{ starts from }\mu\nu\mbox{ and embeds }\theta. \\
	\hspace*{2mm}\Rightarrow
	\rho\mbox{ satisfies }
		(\Phi_\calmdl\cup\{\eta_1 a\neg\eta_2\})\Psi_\calmdl.
	\end{array}\right)
\end{array}
$
\end{center} 
is true. 
}
\\\pf 
By replacing $\eta_1a\eta_2$ with $\eta_1a\neg\eta_2$, 
we can use a proof similar to the one for lemma~\ref{lemma.rsim.se} 
for this lemma.  
\qed

For convenience, 
given two sets $\Delta$ and $\Delta'$ of fairness assumptions,
we let $(\Delta\neg\Delta')$ denote 
\begin{center} 
$\Delta\cup\{\neg\eta\mid \eta\in \Delta'\}\cup
\{\eta_1 a\neg\eta_2\mid \eta_1 a \eta_2\in \Delta'\}$. 
\end{center} 
According to lemmas~\ref{lemma.rsim.ss}, 
\ref{lemma.rsim.se}, \ref{lemma.rsim.ws}, 
and \ref{lemma.rsim.we}, 
we conclude with the following lemma. 

{\lemma \label{lemma.rsim} 
In case $|\Phi_\calspc|+|\Psi_\calspc|\leq 1$, 
a state-pair $\mu\nu$ is not in any simulation of 
$\langle \calmdl,\Phi_\calmdl\Psi_\calmdl\rangle$ 
by $\langle \calspc,\Phi_\calspc\Psi_\calspc\rangle$ iff  
\begin{center}
$\renewcommand{\arraycolsep}{0pt}
\begin{array}{ll}
	& (\mu\mbox{ starts a run $\theta$ of $\calmdl$ satisfying }\Phi_\calmdl\Psi_\calmdl)\\
\wedge	& \forall\rho\left(\begin{array}{l} 
	\rho\mbox{ starts from }\mu\nu\mbox{ and embeds }\theta. \\
	\hspace*{2mm}\Rightarrow
	\rho\mbox{ satisfies }
	  (\Phi_\calmdl\neg\Psi_\calspc)(\Psi_\calmdl\neg\Phi_\calspc).
	\end{array}\right)
\end{array}
$
\end{center} 
is true. 
}
\qed 

A procedure to construct a formula for states $\mu$ that 
starts a run of $\calmdl$ satisfying $\Phi_\calmdl\Psi_\calmdl$ 
can be found in \cite{Wang04a}.   
Lemma~\ref{lemma.rsim} suggests that 
we still need to implement a procedure that 
constructs formulas for state-pairs that start all plays $\rho$ 
satisfying the following constraints. 
\begin{center}
$\forall\rho\left(\begin{array}{l} 
	\rho\mbox{ starts from }\mu\nu\mbox{ and embeds }\theta. \\
	\hspace*{2mm}\Rightarrow
	\rho\mbox{ satisfies }
	  (\Phi_\calmdl\neg\Psi_\calspc)(\Psi_\calmdl\neg\Phi_\calspc).
	\end{array}\right)
$
\end{center} 
Such a play $\rho$ eventually stabilizes into a cycle of state-pairs 
along which 
each assumption in $(\Phi_\calmdl\neg\Psi_\calspc)$ is satisfied 
once and all assumptions in 
$(\Psi_\calmdl\neg\Phi_\calspc)$ are satisfied 
throughout the cycle. 
The following definition characterizes state-pairs in such a cycle.  

{\definition \label{def.isr} 
\underline{\bf CSR}}
A state-pair $\mu\nu$ is {\em CSR} 
({\em Cyclically simulation-refuting}) with 
$(\Phi_\calmdl\neg\Psi_\calspc)(\Psi_\calmdl\neg\Phi_\calspc)$   
iff for every $\phi\in (\Phi_\calmdl\neg\Psi_\calspc)$, 
there exists a run $\theta$ of $\calmdl$ 
with the following two constraints. 
\begin{list1} 
\item[C1:] For every strategy $\sigma$ of $\calspc$ 
	with $\rho=\emplay(\mu\nu,\theta,\sigma)$, 
	if $\rho$ is of infinite length, then the 
	following four constraints are satisfied. 
	\begin{list2} 
	\item[C1a:] All state-pairs along $\rho$ satisfy state-predicates in 
		$(\Psi_\calmdl\neg\Phi_\calspc)$.  
	\item[C1b:] All transition-pairs along $\rho$ satisfy event-predicates in 
		$(\Psi_\calmdl\neg\Phi_\calspc)$.  
	\item[C1c:] For every state-predicate $\eta$ in 
		$(\Phi_\calmdl\neg\Psi_\calspc)$, 
		there is a CSR state-pair in $\rho$ satisfying 
		$\eta$ in 
		more than 1 time units from the start of $\rho$. 
	\item[C1d:] For every event-predicate $\eta$ in 
		$(\Phi_\calmdl\neg\Psi_\calspc)$, 
		there is a transition-pair in $\rho$ satisfying 
		$\eta$ in
		more than 1 time units from the start of $\rho$. 
	\end{list2} 
\item[C2:] There exists a strategy $\sigma$ of $\calspc$ with  
	an infinitely long $\emplay(\mu\nu,\theta,\sigma)$.   
\end{list1} 
The 1-time-unit requirement at condition C1c is for making sure 
that the play is non-Zeno. 

A state-pair $\mu\nu$ is {\em inevitably SR} ({\em ISR}) with 
$(\Phi_\calmdl\neg\Phi_\calspc)(\Psi_\calmdl\neg\Psi_\calspc)$
iff there exists a run $\theta$ of $\calmdl$ from $\mu$ 
such that for all strategies $\sigma$ of $\calspc$, 
if $\emplay(\mu\nu,\theta,\sigma)$ is infinite, 
then $\emplay(\mu\nu,\theta,\sigma)$   
visits a CSR state-pair. 
\qed 

The following lemma is important for our algorithm development. 

{\lemma \label{lemma.simf.neg-evidence} 
Suppose we are given a GBTA $\calmdl$ and a 
BTA $\calspc$.  
For any state-pair 
$\mu\nu\in\bbbbs\langle\calmdl\rangle\times\bbbbs\langle\calspc\rangle$, 
the following two statements are equivalent. 
\begin{list1} 
\item[R1:] $\mu$ starts a run $\theta$ of $\calmdl$ satisfying 
	$\Phi_\calmdl\Psi_\calmdl$ and 
	for all plays $\rho$ from $\mu\nu$ embedding $\theta$, 
	$\rho$ satisfies 
	$(\Phi_\calmdl\neg\Psi_\calspc)(\Psi_\calmdl\neg\Phi_\calspc)$. 
\item[R2:] There exist an $e\in E_\calmdl$, a $t\in \rnneg$, 
	and a $\mu'\in\bbbbs\langle\calmdl\rangle$ 
	with the following constraints. 
	\begin{list2} 
	\item[R2a:] $\mu\stackrel{t,e}{\longrightarrow}\mu'$.   
	\item[R2b:] $\mu'$ starts a run satisfying $\Phi_\calmdl\Psi_\calmdl$.  
	\item[R2c:] For every $f\in E^{(e)}_\calspc$ and 
		$\nu'\in\bbbbs\langle\calspc\rangle$ 
		with $\mu\nu\stackrel{t,ef}\longrightarrow\mu'\nu'$, 
		$\mu'\nu'$ is an ISR state-pair with  
		$(\Phi_\calmdl\neg\Psi_\calspc)
		(\Psi_\calmdl\neg\Phi_\calspc)$. 
	\end{list2} 
\end{list1} 
} 
\pf 
We prove the lemma in two directions.  

\noindent 
$(\Rightarrow)$ 
We assume that R1 is true.  
Conditions R2a and R2b are automatically true since 
$\theta$ must begin with a timed transition step 
$\mu\stackrel{t,e}\longrightarrow\mu'$ for some 
$t\in \rnneg, e\in E_\calmdl$, and $\mu'\in\bbbbs\langle\calmdl\rangle$.  

As for condition R2c, we establish it in the following.  
The truth of R1 means that for every strategy $\sigma$ of $\calspc$,  
if $\rho=\emplay(\mu\nu,\theta,\sigma)$ embeds $\theta$, 
then $\rho$ must satisfy 
$(\Phi_\calmdl\neg\Psi_\calspc)(\Psi_\calmdl\neg\Phi_\calspc)$.  
This means that there exists a $b\in \rnneg$ 
such that for every such infinite $\rho$, 
after $b$ time units from the start of $\rho$, 
all predicates in $(\Psi_\calmdl\neg\Phi_\calspc)$ are satisfied 
and 
all predicates in $(\Phi_\calmdl\neg\Psi_\calspc)$ 
are satisfied infinitely and divergently many times.  
If such a $b$ does not exist, then 
we can construct a play that violates 
$(\Phi_\calmdl\neg\Psi_\calspc)(\Psi_\calmdl\neg\Phi_\calspc)$ 
and the assumption of R1.  
We claim that all state-pairs $\bar\mu\bar\nu$ 
happening after $b$ time units from the start 
in all infinite plays are CSR state-pairs in definition~\ref{def.isr}.  
The reasons are the following. 
\begin{list1} 
\item Since $\bar\mu\bar\nu$ happens $b$ time units after the 
	start of the play, it must satisfy 
	conditions C1a and C1b in definition~\ref{def.isr}.  
	Moreover, along every infinite play from $\bar\mu\bar\nu$, 
	for every predicate $\eta$ in $(\Phi_\calmdl\neg\Psi_\calspc)$, 
	there are infinitely and divergently many state-pairs 
	or transition-pairs that satisfies $\eta$.  
	Thus we can find the first state-pair $\check\mu\check\nu$ in the 
	tail with the following restrictions. 
	\begin{list2} 
	\item $\check\mu\check\nu$ is at least one time unit from 
		the start of the play. 
	\item Either $\check\mu\check\nu$ satisfies $\eta$ 
		as a state-predicate 
		or the transition-pair right before 
		$\check\mu\check\nu$ satisfies $\eta$ 
		as an event-predicate.  
	\end{list2} 
	This implies that conditions C1c and C1d in 
	definition~\ref{def.isr} are satisfied at $\check\mu\check\nu$.  
\item The assumption that leads to the satisfaction of 
	$(\Phi_\calmdl\neg\Psi_\calspc)(\Psi_\calmdl\neg\Phi_\calspc)$
	by $\rho$ then implies that 
	there exists such a play.  
	This implies that condition C2 in definition~\ref{def.isr} 
	is satisfied.  
\end{list1} 
The argument in the above establishes 
that $\bar\mu\bar\nu$ is indeed a CSR state-pair.  
Thus we know that along every infinite play from $\mu'\nu'$, 
we can reach such a $\bar\mu\bar\nu$.   
This implies that $\mu'\nu'$ is an ISR state-pair and 
condition R2c is satisfied.  
Thus the lemma is proven in this direction.  

\noindent 
$(\Leftarrow)$ 
We assume that R2 is true. 
This implies that there exist an $e\in E_\calmdl$, a $t\in \rnneg$, 
and a $\mu'\in\bbbbs\langle\calmdl\rangle$ 
with $\mu\stackrel{t,e}{\longrightarrow}\mu'$ and   
$\mu'$ starting a run satisfying $\Phi_\calmdl\Psi_\calmdl$.  
There are two cases to analyze. 
\begin{list1} 
\item By letting $\theta$ start with $(\mu,\perp,0)(\mu',e,t)$ 
	and followed by the tail from $\mu'$ that satisfies 
	$\Phi_\calmdl\Psi_\calmdl$, 
	we deduce that $\mu$ also starts a run $\theta$ that satisfies 
	$\Phi_\calmdl\Psi_\calmdl$.  
\item Then for all strategies $\sigma$ of $\calspc$ with 
	$\rho=\emplay(\mu\nu,\theta,\sigma)$, 
	we can go to an ISR state-pair $\mu'\nu'$ with 
	$(\Phi_\calmdl\neg\Psi_\calspc)(\Psi_\calmdl\neg\Phi_\calspc)$.  
	This implies that for all infinite plays from $\mu'\nu'$, 
	we can visit a CSR state-pair $\bar\mu\bar\nu$.  
	Then according to the definition of CSR state-pairs, 
	for each predicate $\eta\in(\Phi_\calmdl\neg\Psi_\calspc)$, 
	we can go from $\bar\mu\bar\nu$ along a play 
	with all state-pairs and transition-pairs satisfying the 
	predicates in $(\Psi_\calmdl\neg\Phi_\calspc)$.  
	Moreover, the play visits a CSR state-pair $\check\mu\check\nu$ 
	that either satisfies $\eta$ as a state-predicate 
	or satisfies with the transition-pair immediately 
	before $\check\mu\check\nu$ as an event-predicate.  
	Since $\check\mu\check\nu$ is also CSR, we can then repeat 
	the same argument to fulfill another predicate assumption in 
	$(\Phi_\calmdl\neg\Psi_\calspc)$. 
	By repeating this procedure for all predicates in 
	$(\Phi_\calmdl\neg\Psi_\calspc)$ infinitely many times, 
	we can construct every infinite plays from $\mu\nu$ that 
	embeds $\theta$ mentioned in the last item.  
	This construction then leads to the conclusion that 
	all plays from $\mu\nu$ embedding $\theta$ satisfy 
	$(\Phi_\calmdl\neg\Psi_\calspc)(\Psi_\calmdl\neg\Phi_\calspc)$.
\end{list1} 
This completes the proof of this direction. 	
Since both directions of the proof are done, 
we know the lemma is true. 
\qed

Lemma~\ref{lemma.simf.neg-evidence} suggests the development 
of evaluation algorithm for CSR state-pairs for the solution of 
USF-simulations of GBTAs. 
In the following, we explain how to do this.

\section{A symbolic algorithm for USF-simulation  
  \label{sec.simf.alg}
} 

In this work, we focus on the simulation algorithm 
for a model GBTA by a specification BTA.  
Our algorithm is based on the construction of formulas 
for CSR and ISR state-pairs. 
In the following, we assume the context of a
model GBTA $\langle \calmdl,\Phi_\calmdl\Psi_\calmdl\rangle$ 
and a specification 
BTA $\langle \calspc,\Phi_\calspc\Psi_\calspc\rangle$.  

In subsection~\ref{subsec.tck.mck}, 
we present some symbolic procedures from model-checking 
technology of dense-time systems as our basic building blocks.  
In subsection~\ref{subsec.timed.inevit}, 
we present algorithms for state-pairs that can be forced to 
a goal in one timed transition step.  
In subsection~\ref{subsec.timed.minevit}, 
we use the procedures in subsection~\ref{subsec.timed.inevit} 
to construct a algorithms for state-pairs that can be forced to 
a goal in zero or more timed transition steps. 
In subsection~\ref{subsec.simcheck.alg}, 
we present the algorithm for simulation-checking. 
In subsection~\ref{subsec.simf.complexity}, 
we analyze the complexity of our algorithm.

\subsection{Building blocks from model-checking technology
\label{subsec.tck.mck}
} 

In this subsection, we adapt procedures for TCTL model-checking \cite{ACD90}
for the evaluation of simulation-checking.  

Given a formula $\eta$, 
a run prefix $(\mu_0,e_0,t_0)\ldots(\mu_k,e_k,t_k)$ of $\calmdl$ 
is called an {\em $\eta$-RPrefix} if 
for every $h\in [0,k)$ and $t\in [0,t_{h+1}-t_h]$, 
$\mu_h+t\models \eta$.  
Similarly, a play prefix 
$(\mu_0\nu_0,e_0f_0,t_0)\ldots(\mu_k\nu_k,e_kf_k,t_k)$ of $\calmdl$ 
is called an {\em $\eta$-PPrefix} if 
for every $h\in [0,k)$ and $t\in [0,t_{h+1}-t_h]$, 
$\mu_h\nu_h+t\models \eta$.   

Given a state-pair set $D$, we let 
$\exists\calspc(D)=\{\mu\mid \mu\nu\in D\}$. 
Given a TA $\calspc$ with $P_\calspc=\{p_1,\ldots,p_m\}$ 
and $X_\calspc=\{x_1,\ldots,x_n\}$, 
we let $\exists\calspc(\eta)$ be 
the following formula.
\begin{center} 
$\exists p_1\ldots\exists p_m\exists x_1\ldots\exists x_n
\left(\eta\right)
$.  
\end{center} 
Also given a set $P=\{p_1,\ldots,p_m\}$ and 
a set $X=\{x_1,\ldots,x_n\}$, 
we let $\emass[P,X](\eta)$ be 
the following formula.
\begin{center} 
$\exists p_1\ldots\exists p_m\exists x_1\ldots\exists x_n
\left(\eta\wedge\bigwedge_{x\in X}x=0\right)
$.  
\end{center} 
Standard procedures for constructing state-predicates of existentially 
quantified formulas can be found in 
\cite{HNSY92,Wang03}.  

Given a transition-pair $ef\in E_\calmdl\times E^{(e)}_\calspc$
with $e=(q_1,q_1')$ and $f=(q_2,q'_2)$, 
we let $ef(\eta)$ be the formula of state-pairs 
that may go to state-pairs in $\eta$ through the simultaneous 
execution of $e$ and $f$ respectively.  
Specifically, $ef(\eta)$ is defined as follows. 
\begin{center} 
$\renewcommand{\arraycolsep}{0pt} 
\left(\begin{array}{ll}
	& q_1\wedge q_2\wedge \lambda_\calmdl(q_1)\wedge\lambda_\calspc(q_2) 
\wedge 	
	\tau_\calmdl(e)\wedge\tau_\calspc(f)\\
\wedge 	& \emass[P_\calmdl\cup P_\calspc,X_\calmdl\cup X_\calspc]\left(
	\begin{array}{lll} 
	\eta 
	&\wedge	& \lambda_\calmdl(q'_1)\\
	&\wedge	& \lambda_\calspc(q'_2)
	\end{array}\right)
\end{array}\right)$
\end{center} 
We also need the formulas for the precondition of time-progress
to a state-pair satisfying $\eta_2$ through intermediate 
state-pairs satisfying $\eta_1$. 
Procedures for such formulas can be found in 
\cite{HNSY92,Wang03,Wang08a,Wang08b}.  
We present the formula, denoted $T(\eta_1,\eta_2)$, 
for the readers' convenience in the following. 
\begin{center} 
$\renewcommand{\arraycolsep}{0pt}
\eta_1\wedge\exists t\left(\begin{array}{ll} 
	& t\geq 0\wedge\eta_2+t \\
\wedge 	& \forall t'((t'<t\wedge t'\geq 0)\rightarrow \eta_1+t')
\end{array}\right)
$
\end{center} 
Here $\eta+t$ represents a formula obtained from $\eta$ 
by replacing every clock variable $x$ in $\eta$ with $x+t$.  

We use adapted TCTL formulas 
$\exists \eta_1\until_\calspc\eta_2$ 
in our presentation of the algorithm.  
Specifically, 
$\exists \eta_1\until_\calspc\eta_2$ characterizes those state-pairs 
$\mu\nu$ with the following restrictions. 
\begin{list1} 
\item $\mu\nu$ starts an $\eta_1$-PPrefix $\rho$ that ends at a state-pair 
	satisfying $\eta_2$.   
\item Along the $\rho$ mentioned in the above,  
	all the transitions are of the form $(\perp,f)$ with 
	$f\in E^{(\perp)}_\calspc$.  
\end{list1} 
Following the techniques in \cite{HNSY92,Wang03}, 
we can construct a formula in 
$\bbbbb(P_\calmdl\cup P_\calspc,X_\calmdl\cup X_\calspc)$ 
that characterizes state-pairs satisfying 
$\exists \eta_1\until_\calspc\eta_2$.  
Specifically, the formula is as follows. 
\begin{center} 
$\exists \eta_1\until_\calspc\eta_2\defn
\emlfp Z\left(\eta_2\vee T\left(\eta_1,\bigvee_{f\in E^{(\perp)}_\calspc}
	f(Z)\right)\right) 
$
\end{center} 
Here $\emlfp$ is the least fixpoint operator 
and $\emlfp Z(\beta(Z))$ represents the smallest solution 
to $Z\equiv\beta(Z)$.  

Another type of formulas that we want to use is 
for states $\mu$ of $\calmdl$ that start runs satisfying 
$\Phi_\calmdl\Psi_\calmdl$.  
We denote this formula as $\exists\pfrr\Phi_\calmdl\Psi_\calmdl$ 
for convenience.  
The construction of this formula can be found in \cite{Wang04b}.

\subsection{One-step timed inevitabilities by $\calmdl$ 
\label{subsec.timed.inevit}}

Given a set $D$ of states (or state-pairs), we use 
$\ldabrac D\rdabrac$ to denote a formula that characterizes $D$. 
Given a formula $\eta$, we use $\ldbrac \eta\rdbrac$ to represent 
the set of states (or state-pairs) that satisfies $\eta$. 
Given an $e\in E_\calmdl$, a set $\Psi$ of event weak fairness assumption, 
and a $t\in \rnneg$, 
we use $\langle \calmdl\rangle D_1\nxt^{e\Psi}_t D_2$ to denote 
the set of state-pairs $\mu\nu$ with the following restrictions. 
\begin{list1} 
\item[$M_1$:] There is a $\ldabrac \exists \calspc(D_1)\rdabrac$-RPrefix 
	\begin{center} 
	$(\mu,\perp,t_0)([\mu\stackrel{t,e}\longrightarrow],e,t+t_0)$ 
	\end{center} 
	with the following two restrictions. 
	\begin{list2} 
	\item $[\mu\stackrel{t,e}\longrightarrow]$ 
		is in $\exists\calspc(D_2)$ and 
		satisfies $\exists\pfrr\Phi_\calmdl\Psi_\calmdl$.   
	\end{list2} 
\item[$M_2$:] For every $\ldabrac D_1\rdabrac$-PPrefix 
	\begin{center} 
	$(\mu_0\nu_0,e_0 f_0,t_0)\ldots(\mu_k\nu_k,e_k f_k,t_k)$ 
	\end{center} 
	with 
	\begin{list2} 
	\item $\mu_0\nu_0=\mu\nu$, 
	\item $t_k-t_0=t$, 
	\item $e_k=e$, and 
	\item $\forall h\in [0,k)(e_h=\perp)$, 
	\item For every event weak fairness assumption 
		$\eta_3 a\eta_4\in \Psi$, 
		if $a\in \epsilon_\calmdl(e)$ and $\mu\models \eta_3$, 
		then $[\mu\stackrel{t,e}\longrightarrow]
		\models\eta_4$.
	\end{list2} 
	$\mu_k\nu_k$ is in $D_2$.  
	Note that in the just-mentioned 
	$\ldabrac D_1\rdabrac$-PPrefix, the strategy of 
	$\calspc$ can only use the internal transitions of $\calspc$. 
\end{list1} 
We can use the following TCTL formula to help us characterize 
$\langle \calmdl\rangle D_1\nxt^{e\Psi}_t D_2$.   
Given two state-predicates $\eta_1,\eta_2$, and 
a set $\Psi$ of event formulas for weak fairness assumption, we let 
$\nxt^{e\Psi}_\calmdl(\eta_1,\eta_2)$ be defined as follows. 
\begin{center} 
$\renewcommand{\arraycolsep}{0pt}
\left(\begin{array}{ll} 
	& T\left(\exists\calspc(\eta_1),z=\ccmfs\wedge e\left(
	\begin{array}{ll} 
		& \exists\calspc(\eta_2)\\
	\wedge	& \exists\pfrr\Phi_\calmdl\Psi_\calmdl
	\end{array}\right)\right) 
	\\
\wedge 	& \neg\exists \eta_1\until_\calspc 
	\left(\begin{array}{ll} 
		& z=\ccmfs 
	\wedge	\bigvee_{f\in E^{(e)}_\calspc}ef(\neg\eta_2)\\
	\wedge 	& \bigwedge_{\scriptsize \begin{array}{l} 
		\eta_3 a\eta_4\in \Psi, \\
		e'\in E_\calmdl, \\
		a\in \epsilon_\calmdl(e'),\\
		f'\in E^{(e')}_\calspc
		\end{array}}
		\neg(\eta_3\wedge e'f'(\neg \eta_4))
	\end{array} 
	\right)
\end{array}\right)$
\end{center} 
Here $z$ is an auxiliary clock variable not used in 
$X_\calmdl\cup X_\calspc$.  
The conjunction 
\begin{center} 
$\bigwedge_{\eta_3 a\eta_4\in \Psi, 
		e'\in E_\calmdl, 
		a\in \epsilon_\calmdl(e'),
		f'\in E^{(e')}_\calspc
		}
		\neg(\eta_3\wedge e'f'(\neg \eta_4))$
\end{center} 
in the post-condition is used to 
make sure that no event weak fairness assumptions in 
$\Psi$ is violated.  
It is used to eliminate all state-pairs violating 
an event weak fairness assumption.   
The following lemma shows how to use the above formulas 
to help us evaluating $\langle \calmdl\rangle D_1\nxt^e_t D_2$.  

{\lemma \label{lemma.nxt.form.ext} 
For every $\mu\nu$, $e=(q,q')\in E_\calmdl$, 
$t\in \left[0,\ccmfs\right]$, 
formulas $\eta_1,\eta_2$ of state-pairs, and 
a set $\Psi$ of event weak fairness assumptions, 
$\mu\nu\in \langle\calmdl\rangle 
	\ldbrac\eta_1\rdbrac\nxt^{e\Psi}_t \ldbrac\eta_2\rdbrac$
iff 
\begin{center} 
$\mu\nu\models \exists z(t=\ccmfs-z 
\wedge 	\nxt^{e\Psi}_\calmdl(\eta_1,\eta_2))
$.  
\end{center} 
}
\pf   
We can rewrite condition $M_2$ of $\langle\calmdl\rangle 
	\ldbrac\eta_1\rdbrac\nxt^{e\Psi}_t \ldbrac\eta_2\rdbrac$   
as follows. 
\begin{list1} 
\item[$M_2'$:] There is no $\eta_1$-PPrefix 
	\begin{center} 
	$(\mu_0\nu_0,e_0 f_0,t_0)\ldots(\mu_k\nu_k,e_k f_k,t_k)$ 
	\end{center} 
	with 
	\begin{list2} 
	\item $\mu_0\nu_0=\mu\nu$, 
	\item $t_k-t_0=t$, 
	\item $e_k=e$, 
	\item $\forall h\in [0,k)(e_h=\perp)$, 
	\item $\mu_k\nu_k\not\models\eta_2$, and   
	\item for every $\eta_3 a\eta_4\in \Psi$ 
		and $e'\in E_\calmdl$, $a\in \epsilon_\calmdl(e)$, 
		and $f'\in E^{(e')}_\calspc$, 
		it is not true that 
		$[\mu_{k-1}\nu_{k-1}\stackrel{0,e'f'}]\models \neg\eta_4$. 
	\end{list2} 
\end{list1} 
It is clear that a state-pair satisfies $M_1$ and $M_2$ if and only if 
it satisfies $M_1$ and $M_2'$.  
By renaming $t_0$ as a clock variable $z$ and $t_k$ as constant $\ccmfs$, 
we can use $\ccmfs-z$ to represent $t$.  
This means that $M_1$ and $M_2'$ can be rewritten 
as $z=\ccmfs-t$ and the following two conditions. 
\begin{list1} 
\item[$\check{M}_1$:] There exists an $\exists\calspc(\eta_1)$-RPrefix 
	\begin{center} 
	$(\mu,\perp,z)([\mu\stackrel{\ccmfs-z,e}
		\llrarrow],e,\ccmfs)$ 
	\end{center} 
	with $\calmdl,[\mu\stackrel{\ccmfs-z,e}
		\llrarrow]\models 
	\exists\calspc(\eta_2)\wedge\exists\pfrr\Phi_\calmdl\Psi_\calmdl$.  
\item[$\check{M}_2'$:] There is no $\eta_1$-PPrefix 
	\begin{center} 
	$(\mu_0\nu_0,e_0f_0,t_0)\ldots(\mu_k\nu_k,e_kf_k,t_k)$ 
	\end{center} 
	with $e_k=e$, $\forall h\in[0,k)(e_h=\perp)$, 
	$\mu_0\nu_0=\mu\nu$,  
	$\mu_k\nu_k\models \neg\eta_2$, and    
	for every $\eta_3 a\eta_4\in \Psi$ 
		and $e'\in E_\calmdl$, $a\in \epsilon_\calmdl(e)$, 
		and $f'\in E^{(e')}_\calspc$, 
		it is not true that 
		$[\mu_{k-1}\nu_{k-1}\stackrel{0,e'f'}\longrightarrow]\models \neg\eta_4$. 
\end{list1} 
$\check{M}_1$ means the following.  
\begin{center} 
$\renewcommand{\arraycolsep}{0pt}  
\mu\models T\left(\exists\calspc(\eta_1),z=\ccmfs\wedge e\left(
	\begin{array}{ll} 
		& \exists\calspc\ldbrac\eta_2\rdbrac\\
	\wedge	& \exists\pfrr\Phi_\calmdl\Psi_\calmdl
	\end{array}\right)\right)
$
\end{center} 
$\check{M}_2'$ means the following. 
\begin{center} 
$\renewcommand{\arraycolsep}{0pt}
\mu\nu\models\neg\exists \eta_1\until_\calspc 
	\left(\begin{array}{ll} 
		& z=\ccmfs 
	\wedge	\bigvee_{f\in E^{(e)}_\calspc}ef(\neg\eta_2)\\
	\wedge 	& \bigwedge_{\scriptsize \begin{array}{l} 
		\eta_3 a\eta_4\in \Psi, \\
		e'\in E_\calmdl, \\
		a\in \epsilon_\calmdl(e'),\\
		f'\in E^{(e')}_\calspc
		\end{array}}
		\neg(\eta_3\wedge e'f'(\neg \eta_4))
	\end{array} 
	\right)
$ 
\end{center} 
Combining these two formulas together and 
reduce them with the definition of $\nxt^{e\Psi}_\calmdl(\eta_1,\eta_2)$, 
we find that $\mu\nu$ must satisfy 
$t=\ccmfs-z 
\wedge 	\nxt^{e\Psi}_\calmdl(\eta_1,\eta_2)
$.  
Thus the lemma is proven. 
\qed

Based on lemma~\ref{lemma.nxt.form.ext}, 
we can define the following notations for those state-pairs 
that can be forced into either certain destination 
or a transition of $\calmdl$ that $\calspc$ cannot match.  
Specifically, we let 
\begin{center} 
$\langle \calmdl\rangle D_1\nxt^\Psi D_2
\defn 
\bigcup_{e\in E_\calmdl,t\in\rnneg}\langle \calmdl\rangle D_1\nxt^{e\Psi}_t D_2$.
\end{center} 
Correspondingly, given two formulas $\eta_1$ and $\eta_2$, 
we can construct 
$\nxt^\Psi_\calmdl(\eta_1,\eta_2)$, defined 
as follows. 
\begin{center}
$\nxt^\Psi_\calmdl(\eta_1,\eta_2)\defn\bigvee_{e\in E_\calmdl} 
\exists z\left(\nxt^{e\Psi}_\calmdl\left(\eta_1, \eta_2\right)\right)$. 
\end{center}   
Then according to lemma~\ref{lemma.nxt.form.ext}, 
we can establish the following lemma.

{\lemma \label{lemma.until.form} 
For every 
$\mu\nu\in\bbbbs\langle\calmdl\rangle\times\bbbbs\langle\calspc\rangle$, 
formulas $\eta_1,\eta_2$ of state-pairs, 
and set $\Psi$ of event weak fairness assumptions, 
$\mu\nu\in \langle \calmdl\rangle 
\ldbrac\eta_1\rdbrac\nxt^\Psi\ldbrac\eta_2\rdbrac$ 
iff 
$\mu\nu\models\nxt^\Psi_\calmdl\left(\eta_1, \eta_2\right)$.  
}
\\\pf 
We have the following deduction. 
\begin{center} 
$\renewcommand{\arraycolsep}{0pt}
\begin{array}{ll}
\multicolumn{2}{l}{\mu\nu\in \langle \calmdl\rangle 
	\ldbrac\eta_1\rdbrac\nxt^\Psi\ldbrac\eta_2\rdbrac} \\ 
\equiv 
& \mu\nu\in 
	\bigcup_{e\in E_\calmdl,t\in\rnneg}
	\langle \calmdl\rangle D_1\nxt^{e\Psi}_t D_2 \\
\equiv 
&  \bigvee_{e\in E_\calmdl,t\in\rnneg}
	\mu\nu\in\langle \calmdl\rangle D_1\nxt^{e\Psi}_t D_2
\end{array}$
\end{center} 
According to lemma~\ref{lemma.nxt.form.ext}, 
this implies the following. 
\begin{center} 
$\renewcommand{\arraycolsep}{0pt}
\begin{array}{ll} 
\equiv 	& \bigvee_{e\in E_\calmdl,t\in\rnneg}
	\mu\nu\models  
	\exists z\left(\begin{array}{ll}
		& t=\ccmfs-z \\
	\wedge	& \nxt^{e\Psi}_\calmdl(\eta_1,\eta_2)
	\end{array}\right)
	\\
\equiv & \bigvee_{e\in E_\calmdl}
	\mu\nu\models \bigvee_{t\in \rnneg}
	\exists z\left(\begin{array}{ll}
		& t=\ccmfs-z \\
	\wedge 	& \nxt^{e\Psi}_\calmdl(\eta_1,\eta_2)
	\end{array}\right)\\
\equiv & \bigvee_{e\in E_\calmdl}
	\mu\nu\models 
	\exists z\bigvee_{t\in \rnneg}\left(\begin{array}{ll}
		& t=\ccmfs-z \\
	\wedge 	& \nxt^{e\Psi}_\calmdl(\eta_1,\eta_2)
	\end{array}\right)\\
\end{array}$
\end{center} 
Since $\nxt^{e\Psi}_\calmdl(\eta_1,\eta_2)$ does not contain variable $t$, 
the above formulas are equivalent to the following.  
\begin{center} 
$\renewcommand{\arraycolsep}{0pt}
\begin{array}{ll} 
\equiv & \bigvee_{e\in E_\calmdl}
	\mu\nu\models 
	\exists z\left(\begin{array}{ll}
		& \left(\bigvee_{t\in \rnneg}t=\ccmfs-z\right) \\
	\wedge 	& \nxt^{e\Psi}_\calmdl(\eta_1,\eta_2)
	\end{array}\right)\\
\end{array}$
\end{center} 
Since $\bigvee_{t\in \rnneg}t=\ccmfs-z$ is a tautology, 
we have the following. 
\begin{center} 
$\renewcommand{\arraycolsep}{0pt}
\begin{array}{ll} 
\equiv & \bigvee_{e\in E_\calmdl}
	\mu\nu\models 
	\exists z\left(\nxt^{e\Psi}_\calmdl(\eta_1,\eta_2)
	\right)\\
\equiv & \mu\nu\models\bigvee_{e\in E_\calmdl}
	\exists z\left(\nxt^{e\Psi}_\calmdl(\eta_1,\eta_2)
	\right)\\
\equiv & \mu\nu\models\nxt^\Psi_\calmdl(\eta_1,\eta_2)\\
\end{array}$
\end{center} 
The last step is from the definition of $\nxt^\Psi_\calmdl(\eta_1,\eta_2)$.  
Thus the lemma is proven. 
\qed

Note that before the fulfillment of $\eta_2$,  
$\nxt^\Psi_\calmdl\left(\eta_1, \eta_2\right)$ is satisfied 
with play prefixes with only transitions internal to $\calspc$.

\subsection{Multi-step timed inevitabilities by $\calmdl$ 
\label{subsec.timed.minevit}}

In general, we want to characterize state-pairs from which 
$\calmdl$ can force the fulfillment of $\eta_2$ through 
zero or more timed transition steps of $\calmdl$ that 
do not violate the weak fairness assumptions in $\Psi$.  
We denote the set of such state-pairs as 
$\langle \calmdl\rangle \ldbrac\eta_1\rdbrac\until^\Psi \ldbrac\eta_2\rdbrac$.  
For convenience, given two formulas $\eta_1,\eta_2$ for sets of 
state-pairs, we let 
\begin{center} 
$\until^\Psi_\calmdl(\eta_1, \eta_2)\defn
\emlfp Y\left(\eta_2 
\vee 
\nxt^\Psi_\calmdl(\eta_1, Y)\right)$  
\end{center}
Here $\emlfp$ is the least fixpoint operator. 
$\emlfp Y\left(\eta_2\vee\nxt^\Psi_\calmdl(\eta_1,Y)\right)$ 
specifies a smallest solution to equation 
$Y\equiv \eta_2\vee\nxt^\Psi_\calmdl(\eta_1,Y)$.  
The procedure to construct formulas for such least fixpoints 
can be found in \cite{HNSY92,Wang03}.

{\lemma \label{lemma.until-star.form} 
For every state-pairs $\mu\nu$ and formulas $\eta_1,\eta_2$ for state-pairs, 
$\mu\nu\in \langle\calmdl\rangle 
	\ldbrac\eta_1\rdbrac\until^\Psi \ldbrac\eta_2\rdbrac$
iff 
$\mu\nu\models \until^\Psi_\calmdl(\eta_1, \eta_2)$.  
}
\\\pf 
We can prove this lemma in two directions. 

\noindent  
$(\Rightarrow)$  
We assume that 
$\mu\nu\in \langle\calmdl\rangle 
\ldbrac\eta_1\rdbrac\until^\Psi \ldbrac\eta_2\rdbrac$ 
is true.  
We can prove this by induction on the maximum number $n$ 
of timed transition steps of $\calmdl$ 
to reach state-pairs in $\ldbrac\eta_2\rdbrac$ through 
state-pairs in $\ldbrac\eta_1\rdbrac$.  
In the base case, $n=0$ and $\mu\nu\in\ldbrac\eta_2\rdbrac$.  
Then it is clear that $\mu\nu$ also satisfies 
every formula of the form $\eta_2 \vee \nxt_\calmdl(\eta_1, Y)$.  
Thus, 
$\mu\nu\models \until^\Psi_\calmdl(\eta_1, \eta_2)$ 
in the base case and the lemma is proven. 

Now we assume that this direction of the lemma is true for 
every state-pairs with maximum number no greater than $k$ with 
$k\geq 0$.  
Now we have a state-pair $\mu\nu$ 
with maximum number $k+1$ of timed transition 
steps to reach state-pairs in $\ldbrac\eta_2\rdbrac$ 
through state-pairs in $\ldbrac\eta_1\rdbrac$.  
This implies that there exist 
an $e\in E_\calmdl$ and a $t\in\rnneg$ with 
\begin{center} 
$\mu\nu\models\langle\calmdl\rangle\ldbrac\eta_1\rdbrac\nxt^{e\Psi}_t
\left(\langle \calmdl\rangle\ldbrac\eta_1\rdbrac\until^{\Psi}
\ldbrac\eta_2\rdbrac\right)$.  
\end{center} 
This means that in one timed transition step of $e$ and $t$ time units 
by $\calmdl$, 
we end up in a state-pair $\mu'\nu'$ such that 
within $k$ timed transition of $\calmdl$ steps through state-pairs 
in $\ldbrac\eta_1\rdbrac$, we can go from $\mu'\nu'$ to 
state-pairs in $\ldbrac\eta_2\rdbrac$.  
According to the inductive hypothesis, 
we know that $\mu'\nu'$ satisfies 
$\until^\Psi_\calmdl(\eta_1, \eta_2)$.  
Together, this implies the following deduction. 
\begin{center} 
$\begin{array}{ll} 
\multicolumn{2}{l}{
  \mu\nu\models \nxt^\Psi_\calmdl(\eta_1,\until^\Psi_\calmdl(\eta_1, \eta_2))
}\\
\equiv & 
\mu\nu\models 
\nxt^\Psi_\calmdl(\eta_1,\emlfp Y (\eta_2,\nxt^\Psi_\calmdl(\eta_1, Y))
\end{array}$
\end{center} 
According to the definition of least fixpoint, 
the last step implies 
$\mu\nu\models 
\emlfp Y (\eta_2,\nxt^\Psi_\calmdl(\eta_1, Y))$.  
By definition, this implies that 
$\mu\nu\models 
\until^\Psi_\calmdl(\eta_1, \eta_2)$.  
Thus this direction of the lemma is proven by induction. 

\noindent  
$(\Leftarrow)$  
We assume that there exist 
$Y_0,Y_1,\ldots,Y_n$ such that $Y_0=\eta_2$, 
$Y_n=\eta_2\vee\nxt^\Psi_\calmdl(\eta_1,Y_n)$, and 
for every $i\in [0,n)$, 
$Y_{i+1}=\eta_2\vee\nxt^\Psi_\calmdl(\eta_1,Y_i)$.  
We prove by induction on $k\in [0,n]$ that 
$\mu\nu\models Y_k$ implies 
$\mu\nu\in \langle\calmdl\rangle 
\ldbrac\eta_1\rdbrac\until^\Psi\ldbrac\eta_2\rdbrac$.  
The base case is that $k=0$ and $\mu\nu\models \eta_2$.  
This implies that $\mu\nu\in\ldbrac\eta_2\rdbrac$ and 
$\mu\nu\in\langle\calmdl\rangle 
\ldbrac\eta_1\rdbrac\until^\Psi\ldbrac\eta_2\rdbrac$.  
Thus the base case is proven. 

Now we assume that the lemma in this direction is 
true for all $i\in[0,k]$.  
Now we have a $\mu\nu\models Y_{k+1}$.  
This means that 
$\mu\nu\models \eta_2\vee\nxt^\Psi_\calmdl(\eta_1,Y_k)$.  
There are two cases to analyze.  
The first is $\mu\nu\models\eta_2$ and coincides with the base case.  
Thus the first case is already proven.  

The second case is $\mu\nu\models\nxt^\Psi_\calmdl(\eta_1,Y_k)$.  
According to lemma~\ref{lemma.nxt.form.ext}, 
this implies that we can force in one timed transition step 
through state-pairs in $\ldbrac\eta_1\rdbrac$ 
to state-pairs $\mu'\nu'$ in $\ldbrac Y_k\rdbrac$.  
Moreover, the inductive hypothesis says that all such  
$\mu'\nu'\in\langle \calmdl\rangle\eta_1\until^\Psi\eta_2$.  
According to the definition of $\langle \calmdl\rangle\eta_1\until^\Psi\eta_2$, 
this implies that 
$\mu\nu\in\langle \calmdl\rangle\eta_1\until^\Psi\eta_2$.  
Thus the lemma is proven in this direction. 

Thus the lemma is proven. 
\qed

\subsection{Simulation checking algorithm 
\label{subsec.simcheck.alg} 
} 

Our plan is first to use the procedures in 
subsections~\ref{subsec.tck.mck}, 
\ref{subsec.timed.inevit}, and \ref{subsec.timed.minevit} 
to construct a procedure for evaluating CSR state-pairs.  
Then we use this procedure to evaluate ISR state-pairs.  
For convenience, 
we denote 
\begin{center} 
$\emspms\defn V_\calmdl\wedge V_\calspc
\wedge\left(\bigwedge_{\mbox{\scriptsize state-predicate }
\psi\in(\Psi_\calmdl\neg\Phi_\calspc)}\psi\right)$.  
\end{center} 
Conceptually, $\emspms$ denotes the state-predicates  
that a play satisfying 
$(\Phi_\calmdl\neg\Psi_\calspc)
(\Psi_\calspc\neg\Phi_\calmdl)$ 
must stabilize with. 
Also we let $\emepms$ be the set of event-predicates in 
$(\Psi_\calmdl\neg\Phi_\calspc)$.  
For convenience, we also let 
$\Phi=(\Phi_\calmdl\neg\Psi_\calspc)$ and 
$\Psi=(\Psi_\calspc\neg\Phi_\calmdl)$.  

We present a greatest fixpoint characterization, 
denoted $\emufms(\eta)$, 
of the CSR state-pairs with an MF-assumption $\Phi\Psi$.    
A state-pair $\mu\nu$ satisfies $\emufms(\eta)$ if there is   
a fair run from $\mu$ such that 
all plays embedding the run from $\mu\nu$ cannot be fair for $\calspc$.  
The characterization follows.  
\begin{center} 
$\renewcommand{\arraycolsep}{0pt}
\emufms\defn
\emgfp W.\left(\bigwedge_{\phi\in \Phi}
	\until^{\emsepms}_\calmdl 
		\left(\emspms, \left(W\wedge\phi\right)\right)\right)$.  
\end{center} 
Here $\emgfp$ is the greatest fixpoint operator. 
$\emgfp W.\left(\beta(W)\right)$ is 
a largest solution $W$ to $W\equiv \beta(W)$.  
The procedure to construct formulas for greatest fixpoints 
can be found in \cite{HNSY92,Wang03}.  
The following lemma establishes the correctness of the characterization. 

{\lemma \label{lemma.mfsf} 
A state-pair $\mu\nu$ is CSR with  
$(\Phi_\calmdl\neg\Psi_\calspc)(\Psi_\calspc\neg\Phi_\calmdl)$ 
iff $\mu\nu\models\emufms$.
} 
\\\pf 
Following definition~\ref{def.isr}, 
lemma~\ref{lemma.until-star.form}, 
the definition of $\emspms$, and the semantics of greatest fixpoint, 
$\emufms$ is actually a rewriting of the CSR definition 
with logic formulas, the greatest fixpoint procedure, 
and the $\until^\Psi_\calmdl()$ procedure.
Thus the lemma is proven.  
\qed 

Now we use $\emufms$ to evaluate ISR state-pairs.  
Given a fair run $\theta$ of $\calmdl$, 
there are two classes of ISR state-pairs.   
The first class contains state-pairs that 
start no play embedding $\theta$.    
The second class contains state-pairs 
with a strategy $\calspc$ to drive a play to stabilize to CSR state-pairs. 
The former can be evaluated with the traditional procedures 
for branching simulation \cite{Cerans92,TAKB96,WL97}.  
Specifically, state-pairs is in the first class 
can be characterized with the following lemma. 

{\lemma \label{lemma.isr.1st} 
A state-pair $\mu\nu$ is a first-class ISR state-pair 
iff 
$\renewcommand{\arraycolsep}{0pt}
\mu\nu\models 
\until^\emptyset_\calmdl\left(V_\calmdl\wedge V_\calspc, 
\left(\begin{array}{ll} 
	& T(V_\calmdl\wedge V_\calspc, 
	V_\calmdl\wedge \neg V_\calspc)\\
\vee 	& \bigvee_{e\in E_\calmdl, f\in E^{(e)}_\calspc}
	ef(V_\calmdl\wedge \neg V_\calspc)
\end{array}\right)\right)$.
} 
\\\pf 
$\mu\nu$ is first class iff for all strategies $\sigma$ of $\calspc$, 
$\emplay(\mu\nu,\theta,\sigma)$ is of finite length.  
There can only be two causes for the termination of the plays. 
\begin{list1} 
\item Along a time progress operation, 
	$\calmdl$ moves to a valid state while 
	$\calspc$ cannot.  
	This is captured by formula 
	$T(V_\calmdl\wedge V_\calspc, 
	V_\calmdl\wedge \neg V_\calspc)$.  
\item At a transition $e$ by $\calmdl$, 
	no compatible $f\in E^{(e)}_\calspc$ can result in 
	a valid state of $\calspc$.  
	This is captured by 
	$ef(V_\calmdl\wedge \neg V_\calspc)$.  
\end{list1} 
If and only if 
$\calmdl$ can drive all plays to state-pairs with these two 
causes, 
then it is clear all plays are finite in length. 
Thus the lemma is proven. 
\qed

The state-pairs in the second class can be forced into infinite plays  
that stabilize in CSR state-pairs.  
Specifically, we have the following lemma. 

{\lemma \label{lemma.isr.2nd} 
A state-pair $\mu\nu$ is a second-class ISR state-pair 
iff $\mu\nu\models 
\until^\emptyset_\calmdl\left(V_\calmdl\wedge V_\calspc, \emufms\right)$.
} 
\\\pf 
This lemma follows from the definition of the second class state-pairs, 
lemma~\ref{lemma.until-star.form},  
and lemma~\ref{lemma.mfsf}.  
\qed

Combining 
lemmas~\ref{lemma.simf.neg-evidence}, 
\ref{lemma.isr.1st}, and 
\ref{lemma.isr.2nd}, 
we present the following lemma for the characterization of 
state-pairs that is in no simulation of a GBTA by a BTA. 

{\lemma \label{lemma.sim.refute}
A state-pair $\mu\nu$ is in no simulation of 
a GBTA $(\calmdl,\Phi_\calmdl\Psi_\calmdl)$ by 
a BTA $(\calspc,\Phi_\calspc\Psi_\calspc)$ 
iff 
$\mu\nu$ satisfies either 
$\renewcommand{\arraycolsep}{0pt}
\until^\emptyset_\calmdl\left(V_\calmdl\wedge V_\calspc, 
\left(\begin{array}{ll} 
	& T(V_\calmdl\wedge V_\calspc, 
	V_\calmdl\wedge \neg V_\calspc)\\
\vee 	& \bigvee_{e\in E_\calmdl, f\in E^{(e)}_\calspc}
	ef(V_\calmdl\wedge \neg V_\calspc)
\end{array}\right)\right)$ or 
$\until^\emptyset_\calmdl\left(V_\calmdl\wedge V_\calspc, \emufms\right)$.  
} 
\qed

\subsection{Complexity \label{subsec.simf.complexity}} 

The complexity of our algorithm relies on the implementation of 
the basic manipulation procedures of zones.  
Like in \cite{HNSY92}, we argue that we can implement 
the formulas as sets of pairs of proposition valuations and regions \cite{ACD90}.  
In such an implementation, basic operations like subsumption, 
intersection, union, complement, time progression, and 
variable quantification can all be done in EXPTIME.  

{\lemma \label{lemma.bisim.complexity} 
Proper implementations of the formulation in lemma~\ref{lemma.sim.refute} 
can be done in EXPTIME.  
}
\\\pf 
According to \cite{ACD90}, a zone 
can be implemented as a set of regions.  
The number of regions is exponential to the input size 
of $(\calmdl,\Phi_\calmdl\Psi_\calmdl)$ and 
$(\calspc,\Phi_\calspc\Psi_\calspc)$. 
All precondition calculations need at most polynomial 
numbers of region set operations and 
can all be done in EXPTIME.  
The numbers of 
iterations of the least and greatest fixpoint procedures 
are at most the number of regions.  
Thus, summing everything up, we conclude that our algorithm 
can be executed in EXPTIME. 
\qed

\section{Simulation-checking against a shared environment
\label{sec.sim.env}
}

In real-world, we may usually want 
to check whether a system component satisfies its specification.  
In such a context, the simulation-checking is carried out 
against the same behavior of the environment of the component. 
Such a context can usually make room for verification efficiency 
if we carefully represent the common environment state information.  
In this section, we extend the simulation defined in 
section~\ref{sec.simf} to simulation of a model by a specification 
against a common environment. 
Then we propose a technique to take advantage of the common 
environment information for simulation-checking efficiency.

\begin{figure*}
\begin{center} 
\begin{picture}(0,0)%
\includegraphics{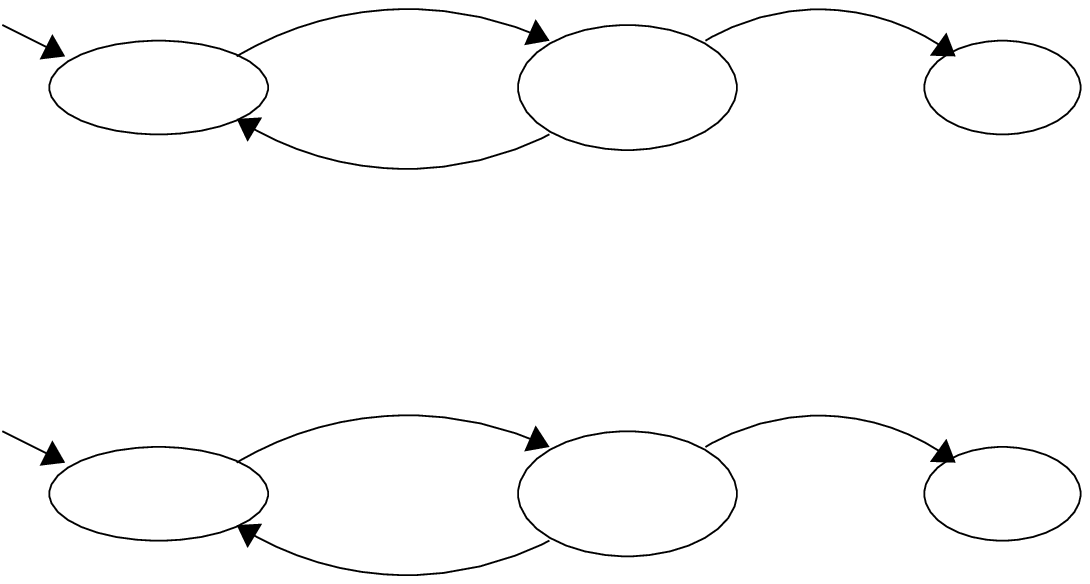}%
\end{picture}%
\setlength{\unitlength}{3947sp}%
\begingroup\makeatletter\ifx\SetFigFontNFSS\undefined%
\gdef\SetFigFontNFSS#1#2#3#4#5{%
  \reset@font\fontsize{#1}{#2pt}%
  \fontfamily{#3}\fontseries{#4}\fontshape{#5}%
  \selectfont}%
\fi\endgroup%
\begin{picture}(5195,3561)(664,-3139)
\put(1501, 89){\makebox(0,0)[lb]{\smash{{\SetFigFontNFSS{12}{14.4}{\rmdefault}{\mddefault}{\updefault}{\color[rgb]{0,0,0}$x_3:=0;$}%
}}}}
\put(2926,-811){\makebox(0,0)[lb]{\smash{{\SetFigFontNFSS{12}{14.4}{\rmdefault}{\mddefault}{\updefault}{\color[rgb]{0,0,0}$!\mbox{\tt serve}$}%
}}}}
\put(5176,-361){\makebox(0,0)[lb]{\smash{{\SetFigFontNFSS{12}{14.4}{\rmdefault}{\mddefault}{\updefault}{\color[rgb]{0,0,0}$\mbox{\tt sleep}$}%
}}}}
\put(3226,-286){\makebox(0,0)[lb]{\smash{{\SetFigFontNFSS{12}{14.4}{\rmdefault}{\mddefault}{\updefault}{\color[rgb]{0,0,0}$\mbox{\tt comp}$}%
}}}}
\put(4726, 89){\makebox(0,0)[lb]{\smash{{\SetFigFontNFSS{12}{14.4}{\rmdefault}{\mddefault}{\updefault}{\color[rgb]{0,0,0}$?\mbox{\tt end}$}%
}}}}
\put(1501,239){\makebox(0,0)[lb]{\smash{{\SetFigFontNFSS{12}{14.4}{\rmdefault}{\mddefault}{\updefault}{\color[rgb]{0,0,0}$?\mbox{\tt request}$}%
}}}}
\put(901,-1111){\makebox(0,0)[lb]{\smash{{\SetFigFontNFSS{12}{14.4}{\rmdefault}{\mddefault}{\updefault}{\color[rgb]{0,0,0}(a) a non-responsive environment process $\calenv$}%
}}}}
\put(976,-361){\makebox(0,0)[lb]{\smash{{\SetFigFontNFSS{12}{14.4}{\rmdefault}{\mddefault}{\updefault}{\color[rgb]{0,0,0}$\mbox{\tt standby}$}%
}}}}
\put(976,-2311){\makebox(0,0)[lb]{\smash{{\SetFigFontNFSS{12}{14.4}{\rmdefault}{\mddefault}{\updefault}{\color[rgb]{0,0,0}$\mbox{\tt standby}$}%
}}}}
\put(1501,-1861){\makebox(0,0)[lb]{\smash{{\SetFigFontNFSS{12}{14.4}{\rmdefault}{\mddefault}{\updefault}{\color[rgb]{0,0,0}$x_3:=0;$}%
}}}}
\put(2926,-2761){\makebox(0,0)[lb]{\smash{{\SetFigFontNFSS{12}{14.4}{\rmdefault}{\mddefault}{\updefault}{\color[rgb]{0,0,0}$!\mbox{\tt serve}$}%
}}}}
\put(5176,-2311){\makebox(0,0)[lb]{\smash{{\SetFigFontNFSS{12}{14.4}{\rmdefault}{\mddefault}{\updefault}{\color[rgb]{0,0,0}$\mbox{\tt sleep}$}%
}}}}
\put(3226,-2236){\makebox(0,0)[lb]{\smash{{\SetFigFontNFSS{12}{14.4}{\rmdefault}{\mddefault}{\updefault}{\color[rgb]{0,0,0}$\mbox{\tt comp}$}%
}}}}
\put(3226,-2386){\makebox(0,0)[lb]{\smash{{\SetFigFontNFSS{12}{14.4}{\rmdefault}{\mddefault}{\updefault}{\color[rgb]{0,0,0}$x_3<10$}%
}}}}
\put(4726,-1861){\makebox(0,0)[lb]{\smash{{\SetFigFontNFSS{12}{14.4}{\rmdefault}{\mddefault}{\updefault}{\color[rgb]{0,0,0}$?\mbox{\tt end}$}%
}}}}
\put(1501,-1711){\makebox(0,0)[lb]{\smash{{\SetFigFontNFSS{12}{14.4}{\rmdefault}{\mddefault}{\updefault}{\color[rgb]{0,0,0}$?\mbox{\tt request}$}%
}}}}
\put(901,-3061){\makebox(0,0)[lb]{\smash{{\SetFigFontNFSS{12}{14.4}{\rmdefault}{\mddefault}{\updefault}{\color[rgb]{0,0,0}(b) a responsive environment process $\calenv$}%
}}}}
\end{picture}%
\end{center} 
\caption{A non-responsive and a responsive environment processes} 
\label{fig.mse} 
\end{figure*}

In figure~\ref{fig.mse}, 
there are two TAs for two environment processes. 
Note that location {\tt comp} in figure~\ref{fig.mse}(b) is labeled 
with a deadline $x_3<10$.  
This means that the environment process in figure~\ref{fig.mse}(b) 
can only stay in location {\tt comp} for at most 10 time units. 
Thus the environment process in figure~\ref{fig.mse}(a) may deliver late 
service while 
the one in figure~\ref{fig.mse}(b) always deliver service in 10 time 
units.  
Against the environment described by figure~\ref{fig.mse}(a), 
the $\calspc$ in figure~\ref{fig.ms_ne}(b) does not simulate 
the $\calmdl$ in figure~\ref{fig.ms_ne}(a) 
since the $\calmdl$ terminates the computation on late service while 
the $\calspc$ never terminates the computation. 
In comparison, 
against figure~\ref{fig.mse}(b), 
the $\calspc$ simulates the $\calmdl$ since the service is always in time.

\subsection{CTA}   

We use {\em CTAs} ({\em communicating timed automata})  
to model the interaction between an environment and 
a model (or a specification). 
The formal definition is in the following. 

{\definition \underline{\bf CTA}
\label{def.cta}} 
A CTA of two TAs $\cala$ and $\calb$, 
in symbols $\cala\times \calb$, 
is a TA with the following constraints. 
\begin{list1} 
\item $P_\calab=P_\cala\cup P_\calb$.  
\item $Q_\calab=Q_\cala\times Q_\calb$. 
\item $\Sigma_\calab=\Sigma_\cala=\Sigma_\calb$. 
\item $X_\calab=X_\cala\cup X_\calb$. 
\item $I_\calab\equiv I_\cala\wedge I_\calb$. 
\item For each $(q_1,q_2)\in Q_\calab$, 
	$\lambda_A((q_1,q_2))
	\equiv \lambda_\cala(q_1)\wedge\lambda_\calb(q_2)$.  
\end{list1} 
For simplicity, we assume that 
$P_\cala\cap P_\calb=\emptyset$,  
$Q_\cala\cap Q_\calb=\emptyset$, and 
$X_\cala\cap X_\calb=\emptyset$.  
Moreover, the transitions of a product TA needs to consider 
the synchronization between the two process TAs.  
Specifically, we let $E_\calab\subseteq E_\cala\times E_\calb$.  
For each $(e,f)\in E_\calab$, one of the following constraints must hold. 
\begin{list1} 
\item $(e,f)$ represents the autonomous execution of 
	a process TA with a transition without any events.  
	Formally speaking, this means 
	at least one of $e$ and $f$ is $\perp$, i.e., 
	no operation. 
	We have the following two cases to explain. 
	\begin{list2} 
	\item If $e\neq\perp$ and $f=\perp$, then 
		$e\in E_\cala$, 
		$\epsilon_\calab((e,f))=\epsilon_\cala(e)=\emptyset$, 
		$\tau_\calab((e,f))=\tau_\cala(e)$, 
		$\pi_\calab((e,f))=\pi_\cala(e)$.  
	\item If $e=\perp$ and $f\neq\perp$, then 
		$f\in E_\calb$, 
		$\epsilon_\calab((e,f))=\epsilon_\calb(f)=\emptyset$, 
		$\tau_\calab((e,f))=\tau_\calb(f)$, 
		$\pi_\calab((e,f))=\pi_\calb(f)$.  
	\end{list2} 
\item $(e,f)$ represents the synchronized execution of the 
	two process TAs respectively with a receiving event 
	and a sending event of the same type. 
	Formally speaking, this means that 
	there is an $a\in \Sigma_\calab$ 
	with the following restrictions. 
	\begin{list2}
	\item Either of the following two is true. 
		\begin{list3} 
		\item $\epsilon_\calab((e,f))=
			\{?a@(\cala), !a@(\calb)\}$, 
			$\epsilon_\cala(e)=\{?a\}$, and  
			$\epsilon_\calb(f)=\{!a\}$.  
		\item $\epsilon_\calab((e,f))=
			\{!a@(\cala), ?a@(\calb)\}$, 
			$\epsilon_\cala(e)=\{!a\}$, and  
			$\epsilon_\calb(f)=\{?a\}$. 
		\end{list3}  
		Note here we blend the process names and the operations 
		into the name of the new events.  
		For example, $?a@(\cala)$ and $!a@(\cala)$ 
		respectively represent the receiving and the sending 
		of event $a$ by process $\cala$.  
	\item $\tau_\calab((e,f))=\tau_\cala(e)\wedge\tau_\calb(f)$. 
	\item $\pi_\calab((e,f))=\pi_\cala(e)\cup\pi_\calb(f)$. 
\qed 
	\end{list2} 
\end{list1}

{\example$\,$ \label{exmp.cta}} 
For the specification $\calspc$ in figure~\ref{fig.ms_ne}(b) 
and the environment $\calenv$ in figure~\ref{fig.mse}(a), 
we have $\calespc$ with attributes in 
table~\ref{tab.cta.attr}.  
\begin{table*}[t]
\begin{center} 
$\begin{array}{rcl} 
P_\calespc & = 
&  \{\mbox{\tt idle}_2, \mbox{\tt wait}_2,\mbox{\tt standby}, 
\mbox{\tt process}, \mbox{\tt sleep}\}\\
Q_\calespc & = 
& \left\{\begin{array}{l}
(\mbox{\tt standby},\mbox{\tt idle}_2), 
(\mbox{\tt standby},\mbox{\tt wait}_2),
(\mbox{\tt process},\mbox{\tt idle}_2), \\
(\mbox{\tt process},\mbox{\tt wait}_2),
(\mbox{\tt sleep},\mbox{\tt idle}_2), 
(\mbox{\tt sleep},\mbox{\tt wait}_2) 
\end{array}\right\} \\ 
X_\calespc & = 
& \{x_2,x_3\} \\ 
I_\calespc & \equiv
& \mbox{\tt idle}_2\wedge x_2=0\wedge\mbox{\tt standby}\wedge x_3=0 \\
\lambda_\calespc & = & 
\left[\begin{array}{l}
(\mbox{\tt standby},\mbox{\tt idle}_2)\mapsto \true, 
(\mbox{\tt standby},\mbox{\tt wait}_2)\mapsto \true,
(\mbox{\tt process},\mbox{\tt idle}_2)\mapsto \true, \\
(\mbox{\tt process},\mbox{\tt wait}_2)\mapsto \true,
(\mbox{\tt sleep},\mbox{\tt idle}_2)\mapsto \true, 
(\mbox{\tt sleep},\mbox{\tt wait}_2)\mapsto \true 
\end{array}\right] \\ 
E_\calespc & = & 
\{((\mbox{\tt standby},\mbox{\tt idle}_2),
    (\mbox{\tt process},\mbox{\tt wait}_2)), 
((\mbox{\tt process},\mbox{\tt wait}_2), 
  (\mbox{\tt standby},\mbox{\tt idle}_2))\} \\ 
\Sigma_\calespc & = & 
\{\mbox{\tt request},\mbox{\tt serve},\mbox{\tt end}\} \\ 
\epsilon_\calespc & = & 
\left[
\begin{array}{l}
((\mbox{\tt standby},\mbox{\tt idle}_2),
    (\mbox{\tt process},\mbox{\tt wait}_2))
    \mapsto\{!\mbox{\tt request}@(\calenv), ?\mbox{\tt request}@(\calspc)\}, \\
((\mbox{\tt process},\mbox{\tt wait}_2), 
  (\mbox{\tt standby},\mbox{\tt idle}_2))
    \mapsto\{?\mbox{\tt serve}@(\calenv), !\mbox{\tt serve}@(\calspc)\}
\end{array}\right] \\ 
\tau_\calespc & = & 
\left[
\begin{array}{l}
((\mbox{\tt standby},\mbox{\tt idle}_2),
    (\mbox{\tt process},\mbox{\tt wait}_2))
    \mapsto x_2>5, \\
((\mbox{\tt process},\mbox{\tt wait}_2), 
  (\mbox{\tt standby},\mbox{\tt idle}_2))
    \mapsto\true
\end{array}\right] \\ 
\pi_\calespc & = & 
\left[
\begin{array}{l}
((\mbox{\tt standby},\mbox{\tt idle}_2),
    (\mbox{\tt process},\mbox{\tt wait}_2))
    \mapsto\{x_3\}, \\
((\mbox{\tt process},\mbox{\tt wait}_2), 
  (\mbox{\tt standby},\mbox{\tt idle}_2))
    \mapsto\{x_2\}
\end{array}\right]
\end{array}$\\[2mm] 
$f=[a_0\mapsto b_0, \ldots, a_k\mapsto b_k,\ldots]$ 
denotes a (partial or total) function $f$ 
with $f(a_0)=b_0, \ldots, f(a_k)=b_k, \ldots$.  
\end{center} 
\caption{Attributes of the CTA of $\calspc$ in figure~\ref{fig.ms_ne}(b) 
and $\calenv$ in figure~\ref{fig.mse}(a).}
\label{tab.cta.attr}
\end{table*} 
\qed 

Since a CTA is also a TA, 
we explain how to interpret the notations about TAs for CTAs.  
Given a state $\alpha$ of $\cala$ and a state $\mu$ of $\calb$, 
$(\alpha,\mu)$ is called a state of $\calab$.   
We say a state $(\alpha,\mu)$ satisfies a state predicate 
$\eta\in \bbbbb(P_\calab,X_\calab)$, 
in symbols $(\alpha,\mu)\models\eta$, with 
the following inductive rules. 
\begin{list1} 
\item For any $p\in P_\cala$, $(\alpha,\mu)\models p$ iff $\alpha\models p$.
\item For any $p\in P_\calb$, $(\alpha,\mu)\models p$ iff $\mu\models p$.
\item For any $x\in X_\cala$, 
	$(\alpha,\mu)\models x\sim c$ iff $\alpha\models x\sim c$.
\item For any $x\in X_\calb$, 
	$(\alpha,\mu)\models x\sim c$ iff $\mu\models x\sim c$.
\item $(\alpha,\mu)\models \neg \eta_1$ iff 
	it is not the case that $(\alpha,\mu)\models \eta_1$. 
\item $(\alpha,\mu)\models \eta_1\vee\eta_2$ iff 
	$(\alpha,\mu)\models \eta_1$ or $(\alpha,\mu)\models \eta_2$. 
\end{list1} 
The state after a transition $(e,f)$ from a state $(\alpha,\mu)$ of CTA, 
denoted $(\alpha,\mu)(e,f)$,  
can also be interpreted as 
$(\alpha e, \mu f)$.  
A timed transition of $t$ time units from a state $(\alpha,\mu)$, 
denoted $(\alpha,\mu)+t$,   
can be defined as $(\alpha+t,\mu+t)$.  
In this way, we can also define the timed transition relation between 
two states $(\alpha,\mu),(\alpha',\mu')$ through  
a transition $(e,f)$ in $t$ time units, 
denoted as  
\begin{center} 
$(\alpha,\mu)\stackrel{t,(e,f)}{\longrightarrow}(\alpha',\mu')$, 
\end{center}   
with the following restrictions. 
\begin{list1} 
\item For all $t'\in [0,t]$, $(\alpha,\mu)+t'\models V_\cala\wedge V_\calb$.  
\item $(\alpha,\mu)+t\models \tau_\cala(e)\wedge\tau_\calb(f)$. 
\item $((\alpha,\mu)+t)(e,f)=(\alpha',\mu')$.  
\end{list1} 
Then a run of $\calab$ can also be defined as a sequence 
\begin{center}
$((\mu_0,\nu_0),(e_0,f_0),t_0)\ldots
((\mu_k,\nu_k),(e_k,f_k),t_k)\ldots$  
\end{center} 
with 
$(\mu_k,\nu_k)\stackrel{t_{k+1}-t_k,(e_{k+1},f_{k+1})}\lllrarrow
(\mu_{k+1},\nu_{k+1})$ for all $k\geq 0$.  

Given a CTA $A\times B$ and an MF-assumption 
$\Phi\Psi$ of $A\times B$, 
$\langle A\times B,\Phi\Psi\rangle$ is called 
a {\em GCBTA} ({\em Generalized communicating BTA}).  
Similarly, 
$\langle A\times B,\Phi\Psi\rangle$ is a {\em CBTA}
({\em Communicating BTA}) if 
$|\Phi|+|\Psi|\leq 1$.

\subsection{Simulation of GCBTAs against an environment} 

{\definition \label{def.simfe} 
{\bf Simulation of GCBTAs against an environment}} 
A {\em simulation} $F^e$ of 
a model GCBTA 
$\langle\calmdl,\Phi_\calmdl\Psi_\calmdl\rangle$ by 
a specification GCBTA 
$\langle\calspc,\Phi_\calspc\Psi_\calspc\rangle$ 
against 
an environment GCBTA 
$\langle\calenv,\Phi_\calenv\Psi_\calenv\rangle$ 
is a binary relation 
$F^e\subseteq 
\bbbbs\langle\calenv\times\calmdl\rangle
\times\bbbbs\langle\calenv\times\calspc\rangle$ 
such that for every $(\alpha,\mu)(\beta,\nu)\in F^e$, 
the following restrictions are satisfied. 
\begin{list1} 
\item[SE1:] $\alpha,\beta\in \bbbbs\langle\calenv\rangle$ with 
	$\alpha=\beta$. 
\item[SE2:] $\mu\in\bbbbs\langle\calmdl\rangle$. 
\item[SE3:] $\nu\in\bbbbs\langle\calspc\rangle$. 
\item[SE4:] For every run $\theta$ of $\calenv\times\calmdl$ 
	from $(\alpha,\mu)$ that satisfies 
	$(\Phi_\calmdl\cup\Phi_\calenv)(\Psi_\calmdl\cup\Psi_\calenv)$,  
	there exists a play $\rho$ from $(\alpha,\mu)(\beta,\nu)$ with the 
	following restrictions. 
	\begin{list2} 
	\item $\rho$ embeds $\theta$ and 
		satisfies 
		$(\Phi_\calenv\cup\Phi_\calmdl\cup\Phi_\calspc)
		(\Psi_\calenv\cup\Psi_\calmdl\cup\Psi_\calspc)$.  
	\item For every transition $(e,f)(e',g)$ along $\rho$, 
		$e=e'\in E_\calenv$.  
	\end{list2} 
\end{list1} 
We say that 
$\langle\calspc,\Phi_\calspc\Psi_\calspc\rangle$ simulates 
$\langle\calmdl,\Phi_\calmdl\Psi_\calmdl\rangle$ against 
environment 
$\langle\calenv,\Phi_\calenv\Psi_\calenv\rangle$,  
in symbols 
\begin{center} 
$\langle\calmdl,\Phi_\calmdl\Psi_\calmdl\rangle
\propto \langle\calspc,\Phi_\calspc\Psi_\calspc\rangle:
\langle\calenv,\Phi_\calenv\Psi_\calenv\rangle$, 
\end{center} 
if there exists a simulation $F^e$ of 
$\langle\calmdl,\Phi_\calmdl\Psi_\calmdl\rangle$ by 
$\langle\calspc,\Phi_\calspc\Psi_\calspc\rangle$ 
against 
$\langle\calenv,\Phi_\calenv\Psi_\calenv\rangle$ 
such that 
for every 
$(\alpha,\mu)\models I_\calenv\wedge V_\calenv\wedge I_\calmdl\wedge V_\calmdl$, 
there exists an 
$(\alpha,\nu)\models I_\calenv\wedge V_\calenv\wedge I_\calspc\wedge V_\calspc$ 
with $(\alpha,\mu)(\alpha,\nu)\in F^e$.  
\qed 

As can be seen, definition~\ref{def.simfe} is more restrictive than 
definition~\ref{def.simf} in their presentations.  
However, we can prove that they are equivalent.  

{\lemma \label{lemma.simfe.eq} 
Given  
an environment GCBTA 
$\langle\calenv,\Phi_\calenv\Psi_\calenv\rangle$,    
a model GCBTA 
$\langle\calmdl,\Phi_\calmdl\Psi_\calmdl\rangle$, and  
a specification GCBTA 
$\langle\calspc,\Phi_\calspc\Psi_\calspc\rangle$, 
$\langle\calenv\times \calmdl,(\Phi_\calenv\cup\Phi_\calmdl)
	(\Psi_\calenv\cup\Psi_\calmdl)\rangle
	\propto
\langle\calenv\times \calspc,(\Phi_\calenv\cup\Phi_\calspc)
	(\Psi_\calenv\cup\Psi_\calspc)\rangle
$ iff  
$\langle\calmdl,\Phi_\calmdl\Psi_\calmdl\rangle \propto
\langle\calspc,\Phi_\calspc\Psi_\calspc\rangle:
\langle\calenv,\Phi_\calenv\Psi_\calenv\rangle$.   
} 
\\\pf 
The backward direction of the proof is straightforward since 
every simulation against an environment in definition~\ref{def.simfe} 
is also 
a simulation in definition~\ref{def.simf}.  
Thus we only have to focus on the forward direction of the proof. 
We first assume that there is 
a simulation $F$ of 
$\langle\calenv\times \calmdl,(\Phi_\calenv\cup\Phi_\calmdl)
	(\Psi_\calenv\cup\Psi_\calmdl)\rangle
$ by 
$\langle\calenv\times \calspc,(\Phi_\calenv\cup\Phi_\calspc)
	(\Psi_\calenv\cup\Psi_\calspc)\rangle
$.  
We can construct $F^e$ as follows. 
\begin{center} 
$F^e\defn\{(\alpha,\mu)(\alpha,\nu)\mid (\alpha,\mu)(\beta,\nu)\in F\}$.  
\end{center} 
Given an $(\alpha,\mu)(\beta,\nu)\in F$, 
it is apparent that $(\alpha,\mu)(\alpha,\nu)$ satisfies 
conditions SE1, SE2, and SE3 of definition~\ref{def.simfe}.  
Then for every runs $\theta$ of $\calenv\times\calmdl$ from 
$(\alpha,\mu)$ satisfying 
$(\Phi_\calmdl\cup\Phi_\calenv)(\Psi_\calmdl\cup\Psi_\calenv)$,  
there exists a play $\rho$ from $(\alpha,\mu)(\beta,\nu)$ such that 
$\rho$ embeds $\theta$ and satisfies 
$(\Phi_\calenv\cup\Phi_\calmdl\cup\Phi_\calspc)
		(\Psi_\calenv\cup\Psi_\calmdl\cup\Psi_\calspc)$.  
Suppose
\begin{center}  
$\begin{array}{ll} 
\rho=	& ((\alpha_0,\mu_0)(\beta_0,\nu_0),(e_0,f_0)(e'_0,g_0),t_0)\\
	& \ldots((\alpha_k,\mu_k)(\beta_k,\nu_k),(e_k,f_k)(e'_k,g_k),t_k)\ldots
\end{array}$
\end{center} 
This implies the following for all $k\geq 0$.  
\begin{center} 
$\begin{array}{rcll} 
\alpha_k & \stackrel{t_{k+1}-t_k,e_k}\longrightarrow 
	& \alpha_{k+1} & \hspace*{10mm} (v1)\\
\mu_k& \stackrel{t_{k+1}-t_k,f_k}\longrightarrow 
	& \mu_{k+1} & \hspace*{10mm} (v2) \\
\beta_k& \stackrel{t_{k+1}-t_k,e'_k}\longrightarrow
	& \beta_{k+1}\\
\nu_k& \stackrel{t_{k+1}-t_k,g_k}\longrightarrow
	& \nu_{k+1} & \hspace*{10mm} (v3) \\
\end{array}$
\end{center} 
Then we can construct a sequence $\rho^e$ as follows. 
\begin{center}  
$\begin{array}{ll} 
\rho^e\defn	& ((\alpha_0,\mu_0)(\alpha_0,\nu_0),(e_0,f_0)(e_0,g_0),t_0)\\
		& \ldots((\alpha_k,\mu_k)(\alpha_k,\nu_k),(e_k,f_k)(e_k,g_k),t_k)\ldots
\end{array}$
\end{center} 
We have the following two claims to prove the lemma. 
\begin{list1} 
\item[CL1:] $\rho^e$ is a play of $(\calenv\times\calmdl)$ by 
	$(\calenv\times\calspc)$ and embeds $\theta$. 
\item[CL2:] $\rho^e$ satisfies 
	$(\Phi_\calenv\cup\Phi_\calmdl\cup\Phi_\calspc)
		(\Psi_\calenv\cup\Psi_\calmdl\cup\Psi_\calspc)$.  
\end{list1} 
Claim CL1 relies on 
the validity that for all $k\geq 0$, 
\begin{center} 
$\begin{array}{rcl} 
(\alpha_k,\mu_k)& \stackrel{t_{k+1}-t_k,(e_k,f_k)}\lllrarrow 
	& (\alpha_{k+1},\mu_{k+1})\\
(\alpha_k,\nu_k)& \stackrel{t_{k+1}-t_k,(e_k,g_k)}\lllrarrow
	& (\alpha_{k+1},\nu_{k+1})
\end{array}$
\end{center} 
These two statements rely on the following three statements. 
\begin{center} 
$\begin{array}{rcll} 
\alpha_k & \stackrel{t_{k+1}-t_k,e_k}\longrightarrow 
	& \alpha_{k+1} & \hspace*{10mm}(v4) \\
\mu_k& \stackrel{t_{k+1}-t_k,f_k}\longrightarrow 
	& \mu_{k+1} & \hspace*{10mm}(v5) \\
\nu_k& \stackrel{t_{k+1}-t_k,g_k}\longrightarrow
	& \nu_{k+1}\\
\end{array}$
\end{center} 
The validity of the above three then follows from 
statements $(v1), (v2)$, and $(v3)$ in the above. 
Thus we know that $\rho^e$ is indeed a play of 
$(\calenv\times\calmdl)\times(\calenv\times\calspc)$.  
Furthermore, the validity of statements $(v4)$ and $(v5)$ 
implies that $\rho^e$ indeed embeds $\theta$.  

Now we want to prove claim CL2.  
For all assumptions in $\Phi_\calenv\cup\Phi_\calmdl$ and 
$\Psi_\calenv\cup\Psi_\calmdl$, 
they are automatically satisfied since 
$\rho^e$ also embeds $\theta$ and 
$\theta$ satisfies $(\Phi_\calenv\cup\Phi_\calmdl)
(\Psi_\calenv\cup\Psi_\calmdl)$.  
For a strong fairness assumption $\phi\in\Phi_\calspc$, 
we have the following two cases to analyze. 
\begin{list1} 
\item {\em $\phi$ is a state-predicate.} 
	We claim that along $\rho^e$, 
	for every $k>0$, there exists 
	an $h>k$ and a $t\in[0, t_{h+1}-t_h]$ with  
	$(\alpha_h,\nu_h)+t\models\phi$.   
	This is true since along $\rho$, 
	$(\alpha_h,\mu_h)(\beta_h,\nu_h)+t\models\phi$ which 
	implies that $\nu_h+t\models\phi$ 
	which in turn implies the claim.  
\item {\em $\phi=\eta_1 a\eta_2$ is an 
	event-predicate.}  
	We claim that along $\rho^e$, 
	for every $k>0$, there exists 
	an $h>k$ with $(\alpha_h,\nu_h)+t_{h+1}-t_h\models\eta_1$, 
	$a\in \epsilon_\calenv(e'_{h+1})\cap\epsilon_\calspc(g_{h+1})$, and 
	$(\alpha_{h+1},\nu_{h+1})\models\eta_2$. 
	This is true since along $\rho$, 
	$(\alpha_h,\mu_h)(\beta_h,\nu_h)+t_{h+1}-t_h\models\eta_1$, 
	$a\in\epsilon_\calmdl(e_{h+1})\cap\epsilon_\calmdl(f_{h+1})\cap
	\epsilon_\calmdl(g_{h+1})$, and 
	 $(\alpha_{h+1},\mu_{h+1})(\beta_{h+1},\nu_{h+})\models\eta_2$.  
	This further implies that
	$\nu_h+t_{h+1}-t_h\models\eta_1$, 
	$a\in\epsilon_\calspc(g_{h+1})$, and 
	 $\nu_{h+1}\models\eta_2$.  
  	In the end, this implies the claim.  
\end{list1} 
For a weak fairness assumption $\psi\in\Psi_\calspc$, 
we have the following two cases to analyze. 
\begin{list1} 
\item {\em $\psi$ is a state-predicate.} 
	We claim that there exists a $k>0$ such that for every  
	$h>k$ and $t\in[0, t_{h+1}-t_h]$, 
	$(\alpha_h,\nu_h)+t\models\psi$.   
	This is true since along $\rho$, 
	$(\alpha_h,\mu_h)(\beta_h,\nu_h)+t\models\psi$ which 
	implies that $\nu_h+t\models\psi$ 
	which in turn implies the claim.  
\item {\em $\psi=\eta_1 a\eta_2$ is an event-predicate.}   
	We claim that along $\rho^e$, 
	there exists a $k>0$ such that for all $h>k$, 
	if $(\alpha_h,\nu_h)+t_{h+1}-t_h\models\eta_1$ and  
	$a\in \epsilon_\calespc((e'_{h+1},g_{h+1}))$, 
	then $(\alpha_{h+1},\nu_{h+1})\models\eta_2$. 
	This is true since along $\rho$, 
	if $(\alpha_h,\mu_h)(\beta_h,\nu_h)+t_{h+1}-t_h\models\eta_1$ and  
	$a\in \epsilon_\calespc((e'_{h+1},g_{h+1}))
	=\epsilon_\calemdl((e_{h+1},f_{h+1}))$, 
	then $(\alpha_{h+1},\mu_{h+1})(\beta_{h+1},\nu_{h+})\models\eta_2$.  
	This further implies that
	if $\nu_h+t_{h+1}-t_h\models\eta_1$ and 
	$a\in \epsilon_\calespc((e_{h+1},g_{h+1}))$, then 
	 $\nu_{h+1}\models\eta_2$.  
  	In the end, this implies the claim.  
\end{list1} 
With the proof of claims CL1 and CL2, 
thus we conclude that the lemma is proven.  
\qed 

According to lemma~\ref{lemma.simfe.eq}, 
we can check the classic simulation in definition~\ref{def.simf} 
by checking the one in definition~\ref{def.simfe}.  
This can be helpful in enhancing the verification performance 
when the common environment between the model and 
the specification is non-trivial.

\subsection{Efficiency techniques 
for simulation against an environment} 

Lemma~\ref{lemma.simfe.eq} implies that 
we can use the following techniques to enhance the 
simulation algorithm against an environment.  
\begin{list1} 
\item Based on condition SE1 of definition~\ref{def.simfe}, 
	we significantly reduce the sizes of the spaces of state-pairs 
	by disregarding state-pairs of the form 
	$(\alpha,\mu)(\beta,\nu)$ with $\alpha\neq\beta$.   
	Since the number of different zones representing $\beta$'s 
	can be exponential to the input size, 
	the reduction can result in exponential speed-up.  
\item By mapping variables in $\beta$ in state-pairs 
	$(\alpha,\mu)(\beta,\nu)$, 
	to those in $\alpha$, 
	we actually only have to record one copy of values for each 
	variables in $\alpha$.  
	Since the size of BDD-like diagrams\cite{Bryant86} is 
	exponential to the number of variables, 
	this technique can also significantly reduce the memory usage in 
	representations with BDD-like diagrams.  
\item In evaluating the precondition of state-pairs, 
	we need to enumerate all the transition pairs of the form 
	$(e,f)(e',g)$ with 
	$e,e'\in E_\calenv$, $f\in E_\calmdl$, and $g\in E_\calspc$.  
	If we use the classic simulation, 
	the enumeration is of size 
	$O(|E_\calenv|^2\cdot|E_\calmdl|\cdot|E_\calspc|)$.  
	But with the simulation against a common environment in 
	definition~\ref{def.simfe}, 
	the enumeration is of size 
	$O(|E_\calenv|\cdot|E_\calmdl|\cdot|E_\calspc|)$.  
	Thus significant reduction in time and space complexity can also 
	be achieved with definition~\ref{def.simfe}.  
\end{list1}

\section{Implementation}  \label{sec.imp}

We have implemented the techniques proposed in this manuscript 
in {\bf RED} 8, a model/simulation-checker for CTAs 
and parametric safety analysis for LHAs based 
on CRD (Clock-Restriction Diagram) \cite{Wang03} and 
HRD (Hybrid-Restriction Diagram) technology \cite{Wang04b}.  
The state-pair spaces are explored in a symbolic on-the-fly style.  
To our knowledge, there is no other tool that supports 
fully automatic simulation checking with GBTAs.  

We used parameterized networks of processes as our benchmarks.  
For a network of $m$ processes, we use integer $1$ through $m$ to 
index the processes.  
Users supply two index lists, 
the first for the indices of the model processes and 
the second for indices of the specification processes.  
The process indices not in the two lists are treated as 
indices of the environment processes.  
For example, we may have a system of 10 processes.  
The following describes a simulation-checking task 
of process 1 (the model) by  
process 1 (the specification).  
\begin{verbatim} 
         1;2; 
\end{verbatim} 
Here processes 3 through 10 are the environment processes.  

To support convenience in presenting fairness assumptions, 
we allow parameterized expressions.  
For example, in table~\ref{tab.sim.reqs}(a), we have a 
simulation requirement with parameterized strong fairness assumptions. 
\begin{table} 
\caption{Two simulation requirements} 
\label{tab.sim.reqs} 
\begin{center} 
(a) One simulation requirement\\\hrule 
{\small 
\begin{verbatim} 
#PS-1 assume {
  strong event {execute@(#PS-1)};   
};  
#PS assume { 
  strong true event {execute@(#PS)} true;  
};  
assume { |k:2..#PS-2, 
  strong true event {execute@(k)}; 
} 
\end{verbatim} 
}
\hrule 
(b) Another simulation requirement \\\hrule 
{\small 
\begin{verbatim} 
#PS-1 
assume {
  strong event {execute@(#PS-1)};   
};  
#PS 
assume { 
  weak idle@(#PS);  
};  
assume { 
  |k:2..#PS-2, 
    strong true event {execute@(k)}; 
} 
\end{verbatim} 
} 
\hrule 
\end{center} 
\end{table}
Here \verb+#PS+ is a parameter for the number of processes.  
Thus for a system of 10 processes, 
process 9 is the model, process 10 is the specification, 
while the others are the environment.  
The last {\tt assume} statement is for the fairness assumption 
of the environment.  
The specification of event-predicates is in the following form. 
\begin{center} 
type [$\eta_1$] a [$\eta_2$]  
\end{center} 
Here type is either `{\tt strong}' or `{\tt weak}.'
[$\eta_1$] and [$\eta_2$] are respectively 
the optional precondition and the optional post-condition.  
We may also use quantified expressions to present 
several fairness assumptions together. 
For example, in the above, 
{\small 
\begin{verbatim} 
    assume { |k:2..#PS-2, 
      strong true event {execute@(k)}; 
    } 
\end{verbatim} 
}
\noindent 
presents the following strong fairness assumptions. 
\begin{verbatim}  
  strong true event {execute@(2)} 
  strong true event {execute@(3)} 
          ...         ... 
  strong true event {execute@(8)} 
\end{verbatim}

\section{Experiments}  \label{sec.exp}

To our knowledge, there is no other tool that supports
fully automatic simulation checking with fairness assumptions 
for TAs as ours.  
So we only experimented with our algorithms.  
We report two experiments. 
The first is for timed branching simulation against 
a common environment without fairness assumptions 
in subsection~\ref{subsec.exp.tsim}.  
Especially, we report the performance enhancement of 
the simulation in definition~\ref{def.simfe} (without 
fairness assumption) 
over the simulation in definition~\ref{def.simf}.  

The second experiment is for simulation against a common  
environment with fairness assumptions 
in subsection~\ref{subsec.exp.fsim}.  
Especially, we use liveness properties in the experiment.  

\subsection{Report of timed branching simulation
\label{subsec.exp.tsim}
} 

We used the following three parameterized benchmarks 
from the literature.  
\begin{list1}
\item[1.] {\em Fischer's timed mutual exclusion algorithm} \cite{Wang03}:
	The algorithm relies on a global lock and a local clock per process
	to control access to a critical section.  
	Two timing constants used are 10 and 19.  
\item[2.] {\em CSMA/CD}\cite{Yovine97}:
	This is the Ethernet bus arbitration protocol with
	collision-and-retry.
	The timing constants used are 26, 52, and 808.
\item[3.] {\em Timed consumer/producer}\cite{SGG04}: 
	There is a buffer, some producers, and some consumers.  
	The producers periodically write data to the buffer if it is empty.  
	The consumers periodically wipe out data, if any, in the buffer. 
	The timing constants used are 5, 10, 15, and 20. 
\end{list1} 
For each benchmark, we use one model process and one specification process.  
All the other processes are environment.  
Also for each benchmark, 
two versions are used, one with a simulation and one without. 
For the versions with a simulation, 
$\calmdl$ and $\calspc$ are identical. 
For the version without, 
$\calmdl$ and $\calspc$ differ in only one process transition or invariance 
condition. 
For example, for the Fischer's benchmark, 
the difference is that the triggering condition of a transition 
to the critical section of $\calspc$ is mistaken. 
The performance data is reported in table~\ref{tab.perf.tsim}.  
\begin{table*}[t]%
\begin{center}
\caption{Performance data of scalability w.r.t. various strategies}
\label{tab.perf.tsim}
\small 
\begin{tabular}{l|l|c||r|r||r|r} 
\hline
		& 		& 	
		& \multicolumn{2}{c||}{Definition~\ref{def.simf}}
		& \multicolumn{2}{c}{Definition~\ref{def.simfe}} \\\cline{4-7} 
benchmarks	& versions 	& $m$	& time & memory & time  & memory  \\ \hline \hline
Fischer's	& Simulation	& 4	& $>1800$s & $>8$M 
					& 31.3s & 320k \\ \cline{3-7}
mutual 		& exists.	& 5	& \multicolumn{2}{c||}{N/A}
					& 92.3s & 664k \\ \cline{3-3}\cline{6-7}
exclusion	& 	 	& 6	& \multicolumn{2}{c||}{ } 
					& 281s & 1319k \\ \cline{2-7} 
($m$		& No 	 	& 4	& $>1800$s & $>8.5$M 
					& 11.7s & 250k  \\ \cline{3-7}
processes 	& simulation	& 5	& \multicolumn{2}{c||}{N/A} 
					& 28.0s & 475k  \\ \cline{3-3}\cline{6-7}
)		& exists.	& 6	& \multicolumn{2}{c||}{ } 
					& 86.7s & 955k \\\hline\hline
CSMA/CD		& Simulation 	& 1	& 0.236s & 102k 
					& 0.098s & 41k 	\\ \cline{3-7}
(1 bus+		& exists. 	& 2	& 72.9s & 1791k 
					& 0.80s & 177k 	\\ \cline{3-7}
$m$ senders	& 	 	& 3	& $>1800$s & $>700$M 
					& 125s & 3503k \\ \cline{2-7} 
)		& No	 	& 1	& 0.144s & 103k  
					& 0.085s & 41k 	\\ \cline{3-7}
		& simulation 	& 2	& 52.9s & 3132k 
					& 2.03s & 203k 	\\ \cline{3-7}
		& exists. 	& 3	& \multicolumn{2}{c||}{N/A} 
					& 25.7s & 2089k \\ \hline\hline  
Consumer \&	& Simulation	& 3	& \multicolumn{2}{c||}{ } 
					& 0.30s & 57k  \\ \cline{3-3}\cline{6-7}
producer	& exists.	& 4	& \multicolumn{2}{c||}{ } 
					& 0.43s & 65k  \\ \cline{3-3}\cline{6-7}
(1 buffer	&  		& 5	& \multicolumn{2}{c||}{N/A} 
					& 0.53s & 75k  \\ \cline{3-3}\cline{6-7} 
+1 producer	& No 		& 3	& \multicolumn{2}{c||}{ } 
					& 0.99s & 70k \\ \cline{3-3}\cline{6-7}
+$m$ consumers	& simulation 	& 4	& \multicolumn{2}{c||}{ } 
					& 1.35s & 775k 	\\ \cline{3-3}\cline{6-7}
)		& exists. 	& 5	& \multicolumn{2}{c||}{ } 
					& 1.16s & 83k	\\\hline 
\end{tabular}
\\        
data collected on a Pentium 4 1.7GHz with 380MB memory running LINUX; \\        
s: seconds; k: kilobytes of memory in data-structure;  
iter'n: the number of iterations\\
\end{center}        
\end{table*}
The CPU time used and the total memory consumption for the data-structures 
in state-space representations are reported.  
As can be seen, the performance of our new simulation 
(definition~\ref{def.simfe}) against 
a common environment is significantly better than 
the classic one (definition~\ref{def.simf}).

\subsection{Report of simulation with fairness assumptions
\label{subsec.exp.fsim}
}

\begin{figure} 
\begin{center}
\begin{picture}(0,0)%
\includegraphics{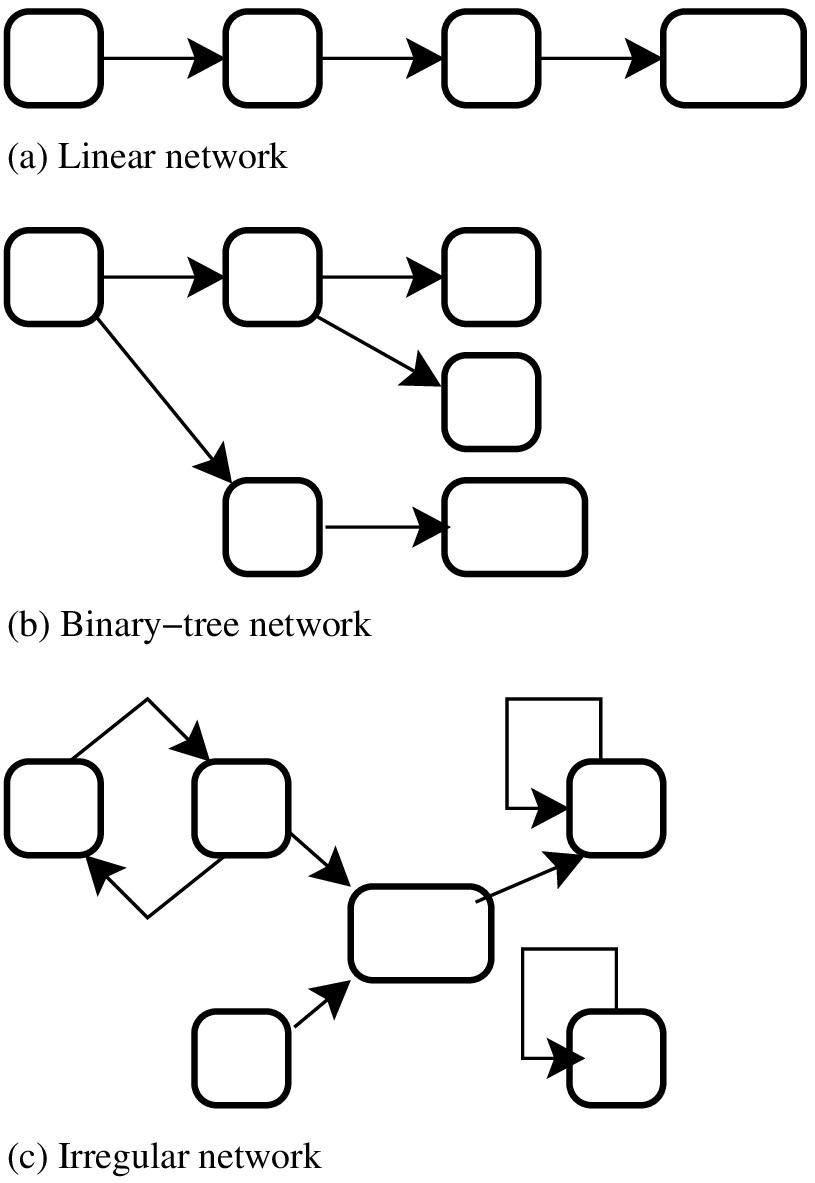}%
\end{picture}%
\setlength{\unitlength}{3947sp}%
\begingroup\makeatletter\ifx\SetFigFontNFSS\undefined%
\gdef\SetFigFontNFSS#1#2#3#4#5{%
  \reset@font\fontsize{#1}{#2pt}%
  \fontfamily{#3}\fontseries{#4}\fontshape{#5}%
  \selectfont}%
\fi\endgroup%
\begin{picture}(3891,5656)(268,-5084)
\put(451,239){\makebox(0,0)[lb]{\smash{{\SetFigFontNFSS{12}{14.4}{\rmdefault}{\mddefault}{\updefault}{\color[rgb]{0,0,0}$A_2$}%
}}}}
\put(1501,239){\makebox(0,0)[lb]{\smash{{\SetFigFontNFSS{12}{14.4}{\rmdefault}{\mddefault}{\updefault}{\color[rgb]{0,0,0}$A_3$}%
}}}}
\put(2551,239){\makebox(0,0)[lb]{\smash{{\SetFigFontNFSS{12}{14.4}{\rmdefault}{\mddefault}{\updefault}{\color[rgb]{0,0,0}$A_4$}%
}}}}
\put(451,-811){\makebox(0,0)[lb]{\smash{{\SetFigFontNFSS{12}{14.4}{\rmdefault}{\mddefault}{\updefault}{\color[rgb]{0,0,0}$A_2$}%
}}}}
\put(1501,-811){\makebox(0,0)[lb]{\smash{{\SetFigFontNFSS{12}{14.4}{\rmdefault}{\mddefault}{\updefault}{\color[rgb]{0,0,0}$A_3$}%
}}}}
\put(1501,-1996){\makebox(0,0)[lb]{\smash{{\SetFigFontNFSS{12}{14.4}{\rmdefault}{\mddefault}{\updefault}{\color[rgb]{0,0,0}$A_4$}%
}}}}
\put(2536,-811){\makebox(0,0)[lb]{\smash{{\SetFigFontNFSS{12}{14.4}{\rmdefault}{\mddefault}{\updefault}{\color[rgb]{0,0,0}$A_5$}%
}}}}
\put(2536,-1411){\makebox(0,0)[lb]{\smash{{\SetFigFontNFSS{12}{14.4}{\rmdefault}{\mddefault}{\updefault}{\color[rgb]{0,0,0}$A_6$}%
}}}}
\put(3526,239){\makebox(0,0)[lb]{\smash{{\SetFigFontNFSS{12}{14.4}{\rmdefault}{\mddefault}{\updefault}{\color[rgb]{0,0,0}$A_5/A_6$}%
}}}}
\put(2476,-2011){\makebox(0,0)[lb]{\smash{{\SetFigFontNFSS{12}{14.4}{\rmdefault}{\mddefault}{\updefault}{\color[rgb]{0,0,0}$A_7/A_8$}%
}}}}
\put(451,-3361){\makebox(0,0)[lb]{\smash{{\SetFigFontNFSS{12}{14.4}{\rmdefault}{\mddefault}{\updefault}{\color[rgb]{0,0,0}$A_2$}%
}}}}
\put(1351,-3361){\makebox(0,0)[lb]{\smash{{\SetFigFontNFSS{12}{14.4}{\rmdefault}{\mddefault}{\updefault}{\color[rgb]{0,0,0}$A_4$}%
}}}}
\put(1351,-4561){\makebox(0,0)[lb]{\smash{{\SetFigFontNFSS{12}{14.4}{\rmdefault}{\mddefault}{\updefault}{\color[rgb]{0,0,0}$A_5$}%
}}}}
\put(3151,-3361){\makebox(0,0)[lb]{\smash{{\SetFigFontNFSS{12}{14.4}{\rmdefault}{\mddefault}{\updefault}{\color[rgb]{0,0,0}$A_3$}%
}}}}
\put(3151,-4561){\makebox(0,0)[lb]{\smash{{\SetFigFontNFSS{12}{14.4}{\rmdefault}{\mddefault}{\updefault}{\color[rgb]{0,0,0}$A_6$}%
}}}}
\put(2026,-3961){\makebox(0,0)[lb]{\smash{{\SetFigFontNFSS{12}{14.4}{\rmdefault}{\mddefault}{\updefault}{\color[rgb]{0,0,0}$A_7/A_8$}%
}}}}
\end{picture}%
\end{center}
\caption{Network topologies of processes}
\label{fig.nets.tops}
\end{figure}

We use a network of TAs as our benchmarks for liveness 
property verification.  
A network consists of $m$ process TAs. 
Process 1 is a dispatcher process. 
Processes 2 through $m-1$ are the environment processes.   
Process $m$ is the model and process $m+1$ is for the specification.  
The execution of a process depends on the incoming services by its 
peer processes.  
In figure~\ref{fig.nets.tops}, 
we draw three example topologies of networks: linear, binary-tree, 
and irregular.  
The nodes represent the processes while the arcs represent 
service channels.  
Inside each node, we put down the name of the TA for the process. 
Note that the model (process $m$) and the specification (process $m+1$) 
have the same channel connections to the other processes. 

The connection relation of the service channels is given 
in a 2-dimensional Boolean array {\em serve}.  
For the linear networks, $\mbox{\em serve}(i,j)$ is true 
iff $i\in [2,m-1]$ and $j=i+1$. 
For the binary-tree networks, 
$\mbox{\em serve}(i,j)$ is true iff $j/2=i$ with integer 
division.  
For the irregular networks, for all $i,j\in[2,m]$, 
$\mbox{\em serve}(i,j)$ is true iff 
$(i * \mbox{\em prime}(i \% 8) + \mbox{\em prime}(j \% 8))$ 
is divisible by 7
where $\mbox{\em prime}(i)$ is the $i$'th prime and 
`$\%$' is the remainder operator.  
For example, in figure~\ref{fig.nets.tops}(c), 
processes 7 and 8, respectively the model and the specification,
are served by both processes 4 and 5. 
Process 6 is only served by itself.  

\begin{figure*}[t] 
\begin{center}
\begin{picture}(0,0)%
\includegraphics{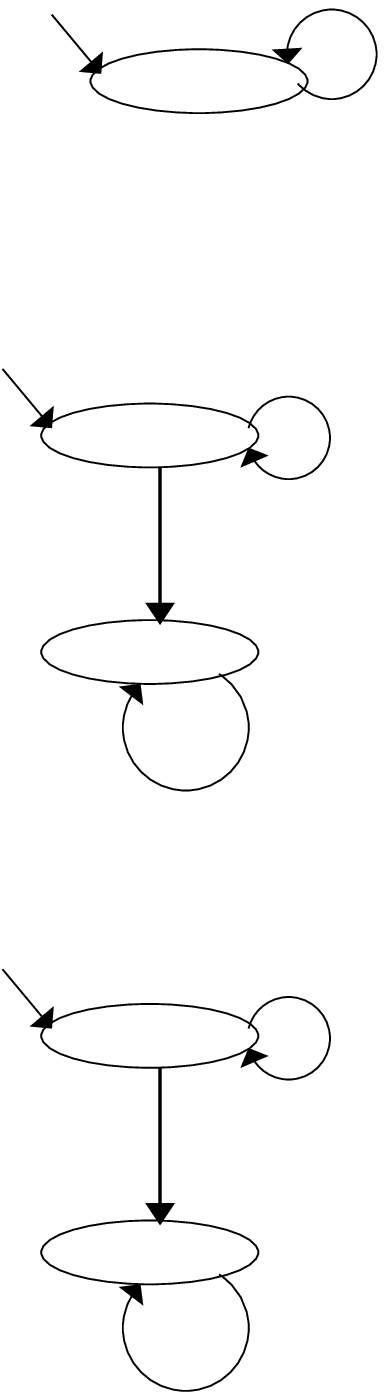}%
\end{picture}%
\setlength{\unitlength}{4144sp}%
\begingroup\makeatletter\ifx\SetFigFontNFSS\undefined%
\gdef\SetFigFontNFSS#1#2#3#4#5{%
  \reset@font\fontsize{#1}{#2pt}%
  \fontfamily{#3}\fontseries{#4}\fontshape{#5}%
  \selectfont}%
\fi\endgroup%
\begin{picture}(6012,6876)(439,-7114)
\put(1171,-871){\makebox(0,0)[lb]{\smash{{\SetFigFontNFSS{12}{14.4}{\rmdefault}{\mddefault}{\updefault}{\color[rgb]{0,0,0}disp}%
}}}}
\put(811,-1456){\makebox(0,0)[lb]{\smash{{\SetFigFontNFSS{12}{14.4}{\rmdefault}{\mddefault}{\updefault}{\color[rgb]{0,0,0}(a) $A_1$ for dispatcher}%
}}}}
\put(1261,-421){\makebox(0,0)[lb]{\smash{{\SetFigFontNFSS{12}{14.4}{\rmdefault}{\mddefault}{\updefault}{\color[rgb]{0,0,0}$!\mbox{\em exec}$}%
}}}}
\put(811,-2491){\makebox(0,0)[lb]{\smash{{\SetFigFontNFSS{12}{14.4}{\rmdefault}{\mddefault}{\updefault}{\color[rgb]{0,0,0}active$_k$}%
}}}}
\put(2026,-2446){\makebox(0,0)[lb]{\smash{{\SetFigFontNFSS{12}{14.4}{\rmdefault}{\mddefault}{\updefault}{\color[rgb]{0,0,0}$x_k:=0;$}%
}}}}
\put(1216,-2896){\makebox(0,0)[lb]{\smash{{\SetFigFontNFSS{12}{14.4}{\rmdefault}{\mddefault}{\updefault}{\color[rgb]{0,0,0}$?\mbox{\em exec}$}%
}}}}
\put(1216,-3256){\makebox(0,0)[lb]{\smash{{\SetFigFontNFSS{12}{14.4}{\rmdefault}{\mddefault}{\updefault}{\color[rgb]{0,0,0}$x_k:=0;$}%
}}}}
\put(1666,-3796){\makebox(0,0)[lb]{\smash{{\SetFigFontNFSS{12}{14.4}{\rmdefault}{\mddefault}{\updefault}{\color[rgb]{0,0,0}$?\mbox{\em exec}$}%
}}}}
\put(991,-4291){\makebox(0,0)[lb]{\smash{{\SetFigFontNFSS{12}{14.4}{\rmdefault}{\mddefault}{\updefault}{\color[rgb]{0,0,0}(b) template for $A_k, k\in[2,m+1]$, requesting service by all incomings}%
}}}}
\put(946,-1996){\makebox(0,0)[lb]{\smash{{\SetFigFontNFSS{12}{14.4}{\rmdefault}{\mddefault}{\updefault}{\color[rgb]{0,0,0}$?\mbox{\em exec}$}%
}}}}
\put(901,-3481){\makebox(0,0)[lb]{\smash{{\SetFigFontNFSS{12}{14.4}{\rmdefault}{\mddefault}{\updefault}{\color[rgb]{0,0,0}idle$_k$}%
}}}}
\put(811,-5236){\makebox(0,0)[lb]{\smash{{\SetFigFontNFSS{12}{14.4}{\rmdefault}{\mddefault}{\updefault}{\color[rgb]{0,0,0}active$_k$}%
}}}}
\put(2026,-5191){\makebox(0,0)[lb]{\smash{{\SetFigFontNFSS{12}{14.4}{\rmdefault}{\mddefault}{\updefault}{\color[rgb]{0,0,0}$x_k:=0;$}%
}}}}
\put(1216,-5641){\makebox(0,0)[lb]{\smash{{\SetFigFontNFSS{12}{14.4}{\rmdefault}{\mddefault}{\updefault}{\color[rgb]{0,0,0}$?\mbox{\em exec}$}%
}}}}
\put(1216,-6001){\makebox(0,0)[lb]{\smash{{\SetFigFontNFSS{12}{14.4}{\rmdefault}{\mddefault}{\updefault}{\color[rgb]{0,0,0}$x_k:=0;$}%
}}}}
\put(1666,-6541){\makebox(0,0)[lb]{\smash{{\SetFigFontNFSS{12}{14.4}{\rmdefault}{\mddefault}{\updefault}{\color[rgb]{0,0,0}$?\mbox{\em exec}$}%
}}}}
\put(991,-7036){\makebox(0,0)[lb]{\smash{{\SetFigFontNFSS{12}{14.4}{\rmdefault}{\mddefault}{\updefault}{\color[rgb]{0,0,0}(c) template for $A_k, k\in[2,m+1]$, requesting service by one incoming}%
}}}}
\put(946,-4741){\makebox(0,0)[lb]{\smash{{\SetFigFontNFSS{12}{14.4}{\rmdefault}{\mddefault}{\updefault}{\color[rgb]{0,0,0}$?\mbox{\em exec}$}%
}}}}
\put(901,-6226){\makebox(0,0)[lb]{\smash{{\SetFigFontNFSS{12}{14.4}{\rmdefault}{\mddefault}{\updefault}{\color[rgb]{0,0,0}idle$_k$}%
}}}}
\put(2026,-2221){\makebox(0,0)[lb]{\smash{{\SetFigFontNFSS{12}{14.4}{\rmdefault}{\mddefault}{\updefault}{\color[rgb]{0,0,0}$x_k>1\wedge \exists h\in[2,m],(\mbox{\em serve}(h,k)\wedge \mbox{\em active}_h)$}%
}}}}
\put(2026,-4966){\makebox(0,0)[lb]{\smash{{\SetFigFontNFSS{12}{14.4}{\rmdefault}{\mddefault}{\updefault}{\color[rgb]{0,0,0}$x_k>1\wedge \left(\begin{array}{ll} & \exists h\in[2,m],\mbox{\em serve}(h,k)\\\wedge&\forall h\in[2,m],(\mbox{\em serve}(h,k)\rightarrow \mbox{\em active}_h)\end{array}\right)$}%
}}}}
\put(1216,-5821){\makebox(0,0)[lb]{\smash{{\SetFigFontNFSS{12}{14.4}{\rmdefault}{\mddefault}{\updefault}{\color[rgb]{0,0,0}$x_k>1\wedge \left(\begin{array}{ll} & \forall h\in[2,m],\neg\mbox{\em serve}(h,k)\\\vee&\exists h\in[2,m],(\mbox{\em serve}(h,k)\wedge \mbox{\em idle}_h)\end{array}\right)$}%
}}}}
\put(6436,-4201){\makebox(0,0)[lb]{\smash{{\SetFigFontNFSS{12}{14.4}{\rmdefault}{\mddefault}{\updefault}{\color[rgb]{0,0,0}$ $}%
}}}}
\put(1216,-3076){\makebox(0,0)[lb]{\smash{{\SetFigFontNFSS{12}{14.4}{\rmdefault}{\mddefault}{\updefault}{\color[rgb]{0,0,0}$x_k>1\wedge \forall h\in[2,m],(\mbox{\em serve}(h,k)\rightarrow \mbox{\em idle}_h)$}%
}}}}
\end{picture}%
\end{center}
\caption{TA templates in a network of $m$ processes}
\label{fig.nets.tas}
\end{figure*}

Templates of the state transition graphs of the processes 
can be found in figure~\ref{fig.nets.tas}.  
Figure~\ref{fig.nets.tas}(a) is the TA for the dispatcher process.  
Specifically, the dispatcher works as a scheduler 
that sends out execution signal, 
\mbox{\em exec},  
to the other processes to allow them to execute.  
There are two templates for the other processes.  
Figures~\ref{fig.nets.tas}(b) and (c) are the two templates for process $k$, 
with $k\in [2,m+1]$, waiting to enter their idle modes. 
A process that uses the template in figure~\ref{fig.nets.tas}(b) 
can execute only when it has received services from all its incoming 
channels.   
A process that uses the template in figure~\ref{fig.nets.tas}(c) 
can execute when it has received services from any of its incoming 
channels.   
Some of the details in notations are 
$P_{A_1}=\emptyset$, $X_{A_1}=\emptyset$, 
$\Sigma_{A_1}=\{\mbox{\em exec}_2,\ldots,\mbox{\em exec}_m\}$, 
and for each $k\in [2,m]$, 
$P_{A_k}=Q_{A_k}$, $\Sigma_{A_k}=\{\mbox{\em exec}_k\}$, and 
	$X_{A_k}=\{x_k\}$.  
Note that in the benchmark, a process may enter the idle mode 
only when all its incoming channels are from idle processes. 
For experiment, we also tried another version of the benchmark 
in which a process may enter the idle mode when any of its incoming channels
is from a idle process.

For each benchmark, we use the two 
simulation requirements in table~\ref{tab.sim.reqs}.  
The performance data is reported in table~\ref{tab.perf.fsim}.  
\begin{table*}[t]%
\begin{center}
\caption{Performance data of scalability w.r.t. various bisimulation definitions}
\label{tab.perf.fsim}
\begin{tabular}{l|c||r|r|r|r|r|r|r|r} 
\hline
\multirow{3}{*}{benchmarks}
& \multirow{3}{*}{$m$} 
& \multicolumn{4}{c|}{service by all incomings} 
& \multicolumn{4}{c}{service by one incoming} \\ \cline{3-10} 
& 
& \multicolumn{2}{c|}{strong} 
& \multicolumn{2}{c|}{weak} 
& \multicolumn{2}{c|}{strong}
& \multicolumn{2}{c}{weak} 
\\ \cline{3-10}
& 
& time  & memory & time & memory 
& time  & memory & time & memory \\ \hline \hline
linear		& 1	& 1.16s & 67k & 0.87s & 67k 
			& 0.44s & 48k & 0.44s & 48k
\\ \cline{2-10}
networks 	& 2	& 1.46s & 122k & 1.88s & 122k 
			& 0.69s & 96k & 0.589s & 97k
\\ \cline{2-10}
	 	& 3	& 2.03s & 191k & 4.22s & 192k 
	 		& 0.93s & 158k & 1.14s & 159k
\\ \cline{2-10} 
		& 4	& 2.60s & 281k 	& 9.70s & 281k 
			& 1.46s & 244k & 1.46s & 244k
\\ \cline{2-10}
		& 5	& 3.46s & 393k & 20.3s & 393k 
			& 1.50s & 359k & 1.48s & 359k
\\ \cline{2-10}
		& 6	& 6.22s & 28.3M & 43.4s & 28.1M 
			& 1.91s & 508k & 2.24s & 508k
\\ \cline{2-10}
		& 7	& 19.7s & 110M & \multicolumn{2}{c|}{N/A} 
			& 3.94s & 26.6M & \multicolumn{2}{c}{N/A} 
\\\hline
tree 		& 1	& 0.94s & 68k	& 0.87s & 68k	
			& 0.41s & 50k 	& 0.41s & 51k
\\ \cline{2-10}
networks	& 2	& 1.23s & 118k	& 1.56s & 119k	
			& 0.72s & 86k	& 0.456s & 87k
\\ \cline{2-10}
	 	& 3	& 1.93s & 194k	& 2.89s & 194k	
	 		& 0.81s & 153k & 0.62s & 153k
\\ \cline{2-10} 
		& 4	& 2.44s & 284k	& 3.89s & 285k	
			& 0.93s & 234k 	& 0.90s & 235k
\\ \cline{2-10}
		& 5	& 3.37s & 412k & 7.18s & 412k 
			& 1.34s & 344k & 1.16s & 345k
\\ \cline{2-10}
 		& 6	& 5.41s & 556k & 10.3s & 557k 
 			& 1.55s & 486k & 1.50s & 487k
\\ \cline{2-10}
 		& 7	& 17.2s & 95.5M & \multicolumn{2}{c|}{N/A} 
 			& 1.98s & 669k & \multicolumn{2}{c}{N/A}
\\ \hline 
general 	& 1	& 1.10s & 105k	& 1.18s & 105k	
			& 0.88s & 191k & 0.91s & 192k
\\ \cline{2-10}
networks	& 2	& 1.06s & 180k	& 0.78s & 180k	
			& 1.47s & 319k 	& 1.15s & 319k
\\ \cline{2-10}
	 	& 3	& 1.15s & 216k	& 0.82s & 216k	
	 		& 1.19s & 343k & 0.92s & 344k
\\ \cline{2-10} 
		& 4	& 1.82s & 436k	& 2.19s & 436k	
			& 2.17s & 947k 	& 3.25s & 947k
\\ \cline{2-10}
		& 5	& 2.06s & 595k & 1.77s & 596k 
			& 2.76s & 1.26M & 2.89s & 1.27M
\\ \cline{2-10}
	 	& 6	& 3.82s & 27.8M & 3.11s & 27.9M 
	 		& 4.92s & 1.56M & 12.8s & 1.56M
\\ \cline{2-10}
	 	& 7	& 16.1s & 107M & \multicolumn{2}{c|}{N/A}
	 		& 16.0s & 90.7M & \multicolumn{2}{c}{N/A}
\\ \hline 
\end{tabular}
\\ 
For each benchmarks, there are a model process, a specification process, 
and $m$ environment processes. `N/A' means ``not avaiable."
\\
data collected on a Pentium 4 1.7GHz with 380MB memory running LINUX; \\        
s: seconds; k: kilobytes of memory in data-structure;  
M: megabytes of total memory \\
\end{center}        
\end{table*}  
As can be seen from the performance data, 
our techniques show promise for the verification of fulfillment of liveness 
properties in concurrent computing.

\section{Concluding remarks } \label{sec.conc}

In this work, we investigate the simulation problem 
of TAs with multiple strong and weak fairness assumptions.  
For the succinct presentation of fairness assumptions, 
we also allow for event fairness properties.  
We then present an algorithm for the USF-simulation of GBTAs.  
The algorithm is based on symbolic model-checking and simulation-checking 
techniques and can be of interest by itself.  
We then propose a new simulation against a common environment 
between the model and the specification.  
We then present efficiency techniques for this new simulation.  
Our implementation and experiment shows the promise 
that our algorithm could be useful in practice in the future.

\section*{Acknowledgment} 

The work is partially supported by NSC, Taiwan, ROC under
grants NSC 97-2221-E-002-129-MY3.\\
Part of the work appears in the proceedings of FORMATS 2007, LNCS 4763, 
Springer-Verlag\cite{Wang07} and 
the proceedings of HSCC 2009, LNCS 5469, Springer-Verlag 
\cite{Wang09a}.

\end{document}